\documentclass[12pt]{article}
\usepackage{graphicx}
\usepackage[utf8]{inputenc}
\usepackage{mathtools}
\usepackage{graphicx,psfrag,epsf,color}
\usepackage{amsmath,amssymb,amsfonts}
\usepackage{array}
\usepackage{cite}
\usepackage{float}
\usepackage{enumitem}
\usepackage{appendix}
\usepackage{comment}
\usepackage{bm}
\usepackage{ascii}
\usepackage[colorlinks]{hyperref}
\usepackage[dvipsnames]{xcolor}
\hypersetup{pageanchor=false,linkcolor=NavyBlue,citecolor=NavyBlue,urlcolor=RoyalPurple}

\renewcommand{\Im}{\textrm{Im}}
\renewcommand{\vec}[1]{\bm{#1}}
\newcommand{\cor}[2]{#1\langle #2 #1\rangle}
\newcommand{\p}{\partial}
\newcommand{\vp}{\varphi}
\newcommand{\ve}{\varepsilon}
\newcommand{\der}{\textrm{d}}
\newcommand{\Lo}{\mathcal{O}}

\newcommand{\Li}{\textrm{Li}}

\numberwithin{equation}{section}

\begin{document}
\allowdisplaybreaks

\begin{titlepage}

\begin{flushright}
{\small
TUM-HEP-1485/23\\
MITP-23-073\\
December 11, 2023 \\
}
\end{flushright}

\vskip1cm
\begin{center}
{\Large \bf Cosmological Correlators in massless $\phi^4$-theory and\\[0.1cm] the Method of Regions}
\end{center}
  \vspace{0.5cm}
\begin{center}
{\sc Martin~Beneke,$^{a}$ \sc Patrick~Hager,$^{b}$ and Andrea~F.~Sanfilippo$^{a}$} 
\\[6mm]
{\it ${}^a$Physik Department T31,\\
James-Franck-Stra\ss e~1,
Technische Universit\"at M\"unchen,\\
D--85748 Garching, Germany}\\[0.2cm]
{\it ${}^b$PRISMA\textsuperscript{+} Cluster of Excellence \& Mainz Institute for Theoretical Physics\\
Johannes Gutenberg University, Staudingerweg 9, D--55128 Mainz, Germany}
\end{center}
\vskip1cm

\begin{abstract}
\noindent 
The calculation of loop corrections to the correlation functions of quantum fields during inflation or in the de~Sitter background presents greater challenges than in flat space due to the more complicated form of the mode functions. While in flat space highly sophisticated approaches to Feynman integrals exist, similar tools still 
remain to be developed for cosmological correlators. However, usually only 
their late-time limit is of interest. We introduce the method-of-region expansion for cosmological correlators as a tool to extract the late-time limit, and illustrate it with several examples for the interacting, massless, minimally coupled scalar field in de~Sitter space. In particular, we consider 
the in-in correlator $\cor{}{\phi^2(\eta,\vec q)\phi(\eta,\vec k_1)\phi(\eta,\vec k_2)}$, whose region structure is relevant to anomalous dimensions and matching coefficients in Soft~de~Sitter effective theory. 
\end{abstract}

\end{titlepage}

\section{Introduction}

\noindent
The inflationary paradigm~\cite{Starobinsky:1980te,Guth:1980zm,Linde:1981mu} 
has revolutionised the modern understanding of the Universe's evolution.
The proposition of an early epoch of quasi-exponential expansion not only elegantly resolves fundamental issues like the flatness and horizon problem but simultaneously provides a mechanism to dilute any potential unobserved relics such as monopoles or domain 
walls. Most importantly for today's observations, the quantum fluctuations of 
fields present during inflation seed the inhomogeneities of the primordial plasma~\cite{Mukhanov:1981xt}, 
which evolve into the inhomogeneity of the cosmic radiation background and 
galaxy distribution.

The behaviour of free quantum fields in curved space-time is well understood, but the intricacies of interacting quantum fields remain an ongoing challenge.
Among these challenges, the study of loop corrections emerges as particularly formidable~\cite{Weinberg:2005vy,Weinberg:2006ac,Senatore:2009cf,Seery:2010kh}.
Nonetheless, incorporating these corrections through a rigorous framework is essential 
to establish internal consistency and to account for the fact that 
infrared (IR) quantum effects in weakly coupled theories can be amplified by 
inflation.

The quantities of interest are the so-called in-in 
correlation functions of the fields \cite{Weinberg:2005vy} at the end of inflation.
Massless 
scalar fields, or those featuring masses much lighter than the Hubble scale, are of 
phenomenological interest as they provide compelling candidates for the 
inflaton field.
During most of 
the inflationary epoch, the space-time can be approximated by the de Sitter space, a maximally-symmetric solution of the Einstein equations describing an exponentially expanding space-time, parameterised by the scale factor $a(t)=e^{H t}$. This offers the possibility of studying the behaviour of interacting 
quantum fields in the early Universe in a simple, well-controlled environment, 
the real, minimally-coupled, massless scalar field with a quartic self-interaction 
$\frac{\kappa}{4!} \phi^4$. Since inflation does not end in exact de Sitter space, 
the relevant observables are the in-in correlators in the late-time 
limit $t\to \infty$.

Despite their apparent simplicity, the computation of late-time (in-in) correlation functions for such scalar fields turns out to be quite difficult. In addition to 
technical difficulties that arise from the far more complicated propagators 
compared to those in flat space-time, the proper choice of regularisation has 
often been a matter of debate. As a consequence, relative to the accumulated 
knowledge on the structure and function space of scattering amplitudes in 
flat space, which has advanced to the three- or even four-loop level, 
the state-of-the art for loop integrals in de Sitter space and curved space-times 
in general is rather modest, even at the one-loop level. 
A major issue arises for light scalar fields due to the well-known infrared 
enhancement, manifested through the emergence of secularly growing logarithmic terms 
proportional to $\ln (a(t))$. These secular terms cause a breakdown of perturbation 
theory for late times $t\to\infty$, as the large logarithms counteract the 
small-coupling suppression of higher-order terms. 

At an even more fundamental level, the free theory of a strictly massless 
scalar field is ill-defined due to the non-existence of a free, de Sitter-invariant vacuum state \cite{Allen:1985ux}, which manifests itself as an IR-divergence 
in the free Wightman function of the scalar field. This difficulty is in fact 
cured in the interacting theory, which self-regularises in the IR. This is 
seen most easily in Euclidean de Sitter space in the path integral formulation, 
where the IR divergence in the free theory arises from a zero mode, which renders 
the path-integral ill-defined  \cite{Rajaraman:2010xd}. 
When turning on the $\frac{\kappa}{4!} \phi^4$ self-interaction, this mode must be 
treated non-perturbatively, and it is possible to develop a well-defined, modified 
perturbative expansion of the Euclidean scalar field correlation functions in 
powers of $\sqrt{\kappa}$ rather than $\kappa$, which does 
not suffer from any IR-divergences \cite{Beneke:2012kn}. It is plausible 
that this structure extends to (Lorentzian) de Sitter space, since 
at least for the massive scalar field, the Euclidean correlators can be 
analytically continued  \cite{Higuchi:2010xt}, and presumably this holds 
for massless $\phi^4$-theory as well \cite{Hollands:2011we}.
However, the secular logarithms in the late-time limit $t\to \infty$ arise only 
upon uncompactifying the time coordinate in the transition to 
de Sitter space, and a separate resummation is required to compute the 
correlators in this limit.

An early solution was provided through Stochastic Inflation 
\cite{Starobinsky:1982ee,Starobinsky:1986fx,Starobinsky:1994bd}, which captures the 
leading infrared dynamics of quantum fields in a fixed de Sitter background beyond 
perturbation theory. The key insight lies in describing the long-wavelength dynamics 
as a stochastic probability distribution governed by a Fokker-Planck equation, where 
the short-range dynamics enter as a Gaussian noise term.
Over time, this stochastic formalism has been rederived through multiple approaches, including the path-integral derivation \cite{Garbrecht:2014dca}, the derivation through the Yang-Feldman equation \cite{Prokopec:2007ak}, a diagrammatic proof \cite{Baumgart:2019clc}, a derivation via the density matrix in the static patch of de Sitter \cite{Mirbabayi:2019qtx}, and the derivation via the Schr\"{o}dinger-formulation of QFT \cite{Collins:2017haz}. While this approach provides an intuitive picture of the evolution of the scalar field, based purely on the stochastic description of the long-wavelength modes, it is unclear how this framework can be systematically extended to include higher-order corrections beyond the Gaussian white-noise approximation, and indeed a generalisation to next-to-leading 
order \cite{Gorbenko:2019rza,Mirbabayi:2020vyt} and next-to-next-to-leading order 
\cite{Cohen:2021fzf} has only recently been achieved in different approaches. 
Since the need for 
resummation arises from the IR and the presence of two scales resulting in the 
small ratio $\frac{k}{a(t) H}\ll 1$, where $k$ denotes a comoving momentum, the natural 
candidate for such a systematic framework is provided by the 
recently developed ``Soft de Sitter effective field theory'' (SdSET) 
\cite{Cohen:2020php,Cohen:2021fzf}.

This brief summary highlights the fact that 
the computation of cosmological correlators 
is always a two-scale problem due to the late-time limit implied, and that 
the computational methods are still in their infancy. In this paper, we point out that the ``method of regions''~\cite{Beneke:1997zp} 
can be employed as a powerful tool to construct the late-time expansion 
of in-in correlation functions in the loop expansion in $\kappa$. The key advantage 
of the method is that the expansion can be constructed without having to 
calculate the expression of the full result before expansion, resulting 
in simpler loop integrals with fewer scales. The method was originally invented 
for the expansion of scattering amplitudes in the non-relativistic 
limit and has since applied to many other kinematic situations. Although fixed-order 
loop integrals 
in massless, minimally coupled $\phi^4$-theory in de Sitter space  may be IR divergent and not well-defined as $t\to \infty$, 
their calculation provides the starting point for the extraction of the 
hard matching coefficients that enter the derivation of the generalised 
stochastic Fokker-Planck equation in the SdSET framework.\footnote{A flat space 
analogue of this statement is the computation of on-shell quark/gluon 
correlation functions in QCD, which are IR divergent. Nevertheless, they 
are the objects from which the hard-matching coefficients of effective 
theories of QCD are computed, since the IR sensitivity drops out in the 
matching. The method of regions provides a tool to extract the matching 
coefficients from the hard region, without the need to compute the full 
correlation function.}

In this paper we introduce the method of regions for cosmological correlators 
of the minimally coupled, massless scalar field in de Sitter space. We work 
through a few examples and analyse the regions that are necessary to 
obtain the expansion of momentum-space correlators in 
\begin{equation}
    -k_i\eta\ll1\,,
\end{equation}
where the external momenta $k_i$ and the (conformal) correlation time $\eta$ 
play the roles of widely separated external scales. We emphasise that the 
region expansion allows one to obtain not only the pole terms, which are 
related to secular and other logarithms, but also the ``finite terms'', 
which are related to the matching coefficients relevant to the stochastic 
approach in higher orders. A peculiar feature of cosmological correlators is 
that already at tree level they present a non-trivial region structure, as 
each vertex is accompanied  by a time integral which must be factorised.
From this point of view, the tree-level in-in correlation functions in de 
Sitter space resemble momentum-space loop amplitudes in flat space. In the language of 
SdSET, this corresponds to the matching 
of ``non-Gaussian initial conditions"~\cite{Cohen:2020php}.\footnote{A 
detailed discussion of the relation of the method-of-region approach 
to SdSET and matching coefficients will be presented in a forth-coming 
paper.} The present work can also be seen in the context of other recent work 
(for example \cite{Arkani-Hamed:2015bza,Arkani-Hamed:2018kmz,Sleight:2019mgd,Melville:2021lst,Baumann:2021fxj, DiPietro:2021sjt,Heckelbacher:2020nue,Heckelbacher:2022hbq,Benincasa:2022gtd}) addressing the analytic structure and computation of cosmological 
correlators for the interacting (not necessarily massless and often 
conformally coupled) scalar field.

The outline of this article is as follows:
in Section~\ref{sec:setup}, we set up the notation and conventions for the treatment of the massless scalar field in de Sitter space. In Section~\ref{sec::regions}, we introduce the method of regions and exemplify its use for de Sitter computations in Section~\ref{treetrispec} for the time-integral that appears in the tree-level trispectrum. In 
Section~\ref{powerspec} the one-loop correction to the power spectrum (two-point function) is presented. With the result from the previous section, this is a very simple calculation, which mainly serves to discuss the issue of mass renormalisation required to keep the scalar field massless. 
The main results and computations of this work refer to the correlator 
$\cor{}{\phi^2(\eta,\vec q)\phi(\eta,\vec k_1)\phi(\eta,\vec k_2)}$, 
contained in Section~\ref{sec::OneLoopMixing}. In this case the full one-loop 
computation appears unnecessarily complicated, and the expansion by regions not only offers a computational tool, but also provides insight into the physics involved.
We conclude in Section~\ref{sec:conclusion}.
Conventions regarding the Schwinger-Keldysh formalism and some further details on computations are summarised in four appendices.

\section{Massless, minimally-coupled scalar field}
\label{sec:setup}

\subsection{Setup and mode functions}
To set the stage, the basic notation and conventions used throughout the article are established.
The background geometry is a four-dimensional de Sitter space parametrised by planar coordinates with line-element
\begin{equation}
\der s^2\equiv g_{\mu\nu}(x)\der x^{\mu}\der x^{\nu}=\der t^2-a(t)^2\der\vec x^2\,,
\end{equation}
employing the metric signature $(+,-,-,-)$.
Here, the scale factor is defined as $a(t) = e^{Ht}$ with the constant Hubble parameter $H$.
For the computations, it is convenient to also introduce the conformal time variable $\eta\in (-\infty,0)$ via the relation 
\begin{equation}
    a(\eta) \equiv -\frac{1}{H\eta}\,.
\end{equation}
In the following, we are interested in correlation functions of the massless, minimally-coupled, self-interacting real scalar field $\phi$, described by the action
\begin{align}
    S &=\int\der^4x\;\sqrt{-g}\biggl[\frac{1}{2}g^{\mu\nu}\p_{\mu}\phi\p_{\nu}\phi-\frac{\kappa}{4!}\phi^4\biggr]\nonumber\\
    &=\int\frac{\der\eta\,\der^3\vec{x}}{(-H\eta)^4}\;\biggl[\frac{(-H\eta)^2}{2}\Bigl[(\p_{\eta}\phi)^2-\p_i\phi\p_i\phi\Bigr] -\frac{\kappa}{4!}\phi^4\biggr]\,,
    \label{fullaction}
\end{align}
where $\kappa$ denotes the coupling constant of the interaction.\footnote{In the gravitational context, $\kappa$ is often reserved for the gravitational coupling constant. Since we are working in a fixed background geometry without dynamic metric perturbations, there is no risk of confusion.} The vacuum state is chosen as the Bunch-Davies vacuum~\cite{Bunch:1978yq} with mode functions
\begin{equation}
    \phi_{\vec k}(\eta)=\frac{\sqrt{\pi}H(-\eta)^{\frac{3}{2}}}{2}H^{(1)}_{\nu}(-k\eta), 
    \qquad k\equiv|\vec{k}|
\label{freemode}
\end{equation}
expressed in terms of the Hankel function of the first kind
\begin{equation}
    H^{(1)}_{\nu}(z)\equiv J_{\nu}(z)+iY_{\nu}(z)\,,
\end{equation}
and the parameter $\nu$ related to the mass $m$ via
\begin{equation}\label{eq:NuParameter}
    \nu\equiv\sqrt{\frac{9}{4}-\frac{m^2}{H^2}}\,.
\end{equation}
For $m\neq0$, this vacuum state is the unique de Sitter-invariant vacuum state that reduces to the Minkowski one in the infinite past $\eta\rightarrow-\infty$, as the modes blue-shift and no longer feel the curvature of space-time.
In the massless case, $\nu=3/2$, and the corresponding Hankel function reads
\begin{equation}\label{eq:MasslessHankelFunction}
    H^{(1)}_{3/2}(z)=-\sqrt{\frac{2}{\pi}}\frac{e^{iz}(i+z)}{z^{3/2}}\,.
\end{equation}

\subsection{Regularisation and Renormalisation}

A suitable regularisation scheme to control the divergences appearing in the loop integrals is dimensional regularisation by analytic continuation to $d=4-2\varepsilon$ dimensional de Sitter space.
It is an analytic regulator, i.e.~it does not introduce a scale into loop integrals, which is required for using the method of regions, as explained below in Sec.~\ref{sec::regions}.
Additionally, dimensional regularisation preserves the full set of de Sitter isometries, namely spatial translations and rotations, dilatations, and special conformal transformations.
In the conformally flat chart, they are generated by the Killing vector fields $P_i$, $R^i_{\;j}$, $D$ and $K^i$ \cite{Arkani-Hamed:2018kmz} 
\begin{equation}
\begin{aligned}
    P_i&=\p_i\,,\label{killing1} &
    R^i_{\;j}&=x^i\p_j-x_j\p^i\,,\\
    D&=-\eta\p_{\eta}-x^i\p_i\,, &
K^i&=2x^i\eta\p_\eta+2x^ix^j\p_j+(\eta^2-\vec x^2)\p_i\,,
\end{aligned}
\end{equation}
in the given order. 
One way to see this is to compute the position-space Wightman function for the scalar field for generic $d$ in this regularisation scheme, which reads
\begin{equation}
    \langle\phi(\eta,\vec{x})\phi(\eta',\vec x')\rangle=\frac{H^{d-2}}{(4\pi)^{\frac{d}{2}}}\frac{\Gamma(\frac{d}{2}-2)\Gamma(\frac{d}{2}+1)}{\Gamma(\frac{d}{2})}\,{_2F_1}\biggl(\frac{d}{2}-2,\frac{d}{2}+1;\frac{d}{2};\frac{1-Z(\eta,\vec x;\eta',\vec x')}{2}\biggr)\,,
\end{equation}
with the de Sitter-invariant distance function
\begin{equation}
    Z(\eta,\vec x;\eta',\vec x')\equiv-1-\frac{(\eta-\eta')^2-|\vec x-\vec x'|^2}{2\eta\eta'}\,,
\end{equation}
which vanishes when acted upon with \eqref{killing1}. Since every diagram can be written in position space as products of the above Wightman function for different arguments, integrated over the positions of the internal vertices using a de Sitter-invariant integration measure, one can conclude that the resulting quantity is de Sitter-invariant as well. 

Compared to flat space, where the mode functions have a unique extension to $d$ dimensions, de Sitter space offers a choice as the $d$-dimensional mode functions depend on the two parameters $d$ and $\nu$ via
\begin{equation}\label{eq::ddimensionalModeFunction}
    \phi_{\vec k}(\eta)=\sqrt{\frac{\pi}{4H}}\,(-H\eta)^{\frac{d-1}{2}}H^{(1)}_{\nu}(-k\eta)\,,\quad \nu = \sqrt{\Bigl(\frac{d-1}{2}\Bigr)^2-\Bigl(\frac{m}{H}\Bigr)^2}\,.
\end{equation}
One convenient choice is to extend the theory to $d$-dimensions by a $d$-dependent mass term of $\Lo(\ve)$
\begin{equation}\label{eq:dDimensionalMassTerm}
    m_d^2 = \frac{H^2(d-4)(d+2)}{4}\,,
\end{equation}
such that the parameter $\nu=3/2$ for any $d$~\cite{Melville:2021lst}.
This ensures that one can always employ the (simpler) Hankel functions~\eqref{eq:MasslessHankelFunction} of the four-dimensional massless scalar field.

To renormalise the theory, one starts from the bare, $d$-dimensional action
\begin{equation}
    S=\int\frac{\der\eta\,\der^{d-1}\vec{x}}{(-H\eta)^d}\;\biggl[\frac{(-H\eta)^2}{2}\Bigl[(\p_{\eta}\phi_0)^2-\p_i\phi_0\p_i\phi_0\Bigr]-\frac{1}{2}m^2_0\phi_0^2-\frac{\kappa_0}{4!}\phi_0^4\biggr]\,,
\end{equation}
which now features a bare mass $m_0$,\footnote{It turns out that the bare mass is proportional to $H^2$. As such, it may also be regarded as a bare non-minimal coupling to the background metric.
In the present context, however, there is no distinction between a bare non-minimal coupling and a bare non-vanishing mass, so it is convenient to treat this as a mass term in the following.} and defines the renormalised quantities as
\begin{equation}
    \phi_0(x)=\sqrt{Z_{\phi}}\,\phi(x)\,,\quad m^2_0=m^2_d+\delta m^2\,,\quad \kappa_0=\tilde{\mu}^{4-d}Z_{\kappa}\kappa\,.
\end{equation}
Here, $\tilde{\mu}=\mu\sqrt{e^{\gamma_E}/4\pi}$ is the standard 't Hooft-scale of the $\overline{\rm MS}$ scheme which renders the renormalised $d$-dimensional coupling $\kappa$ dimensionless, and the $d$-dependent mass term is given in~\eqref{eq:dDimensionalMassTerm}.

When expressed through the renormalised quantities, the action reads
\begin{align}
    S &= \int\frac{\der\eta\,\der^{d-1}\vec{x}}{(-H\eta)^d}\;\biggl[\frac{(-H\eta)^2}{2}\Bigl[(\p_{\eta}\phi)^2-\p_i\phi\p_i\phi\Bigr]-\frac 12 m^2_d\phi^2-\frac{\kappa\tilde{\mu}^{4-d}}{4!}\phi^4\\
        &\phantom{\int\frac{\der\eta\,\der^{d-1}\vec{x}}{(-H\eta)^d}}+\delta_Z\frac{(-H\eta)^2}{2}\Bigl[(\p_{\eta}\phi)^2-\p_i\phi\p_i\phi\Bigr]
    -\frac{1}{2}\delta_m\phi^2
    -\delta_\kappa\frac{\kappa\tilde{\mu}^{4-d}}{4!}\phi^4\biggr]\,,\nonumber
\end{align}
with the counterterms given by
\begin{equation}
    \delta_Z \equiv Z_\phi-1\,,\quad \delta_m \equiv (Z_\phi-1)m_d^2 + Z_\phi\delta m^2\,,\quad \delta_\kappa \equiv Z_\kappa Z_\phi^2-1\,.
\end{equation}
This is the starting point for renormalised perturbation theory. After computing correlation functions in $d$ dimensions, one sets $d=4-2\ve$ and expands the result for small $\ve$. The counterterms $\delta_i$ are then computed in the standard way as a power series in $\kappa$. 
For the following considerations, it is sufficient to work with the tree-level coupling $\kappa$, setting $Z_\kappa=1$.
In addition, the wave-function renormalisation starts at two-loop like in flat space, $Z_\phi = 1 + \mathcal{O}(\kappa^2)$, and its contribution can be neglected for the one-loop discussion.
The only remaining counterterm is the mass counterterm $\delta_m = \delta m^2$.
The corresponding renormalisation scheme is discussed in Sec.~\ref{full1loop} when the power spectrum is computed.

\subsection{In-in/Schwinger-Keldysh formalism}
The observables of interest, the in-in correlation functions, are computed in the so-called in-in or Schwinger-Keldysh formalism \cite{Schwinger,Keldysh}. A summary of the setup and the momentum-space diagrammatic rules of this formalism are presented in App.~\ref{SKrules}. The fundamental quantity for the following computations is the momentum-space Wightman function for the $d$-dimensional renormalised field
\begin{equation}
\cor{}{\phi(\eta,\vec k)\phi(\eta',\vec k')}=(2\pi)^{d-1}\delta^{({d-1})}(\vec k+\vec k')\frac{\pi}{4H}(-H\eta)^{\frac{d-1}{2}}(-H\eta')^{\frac{d-1}{2}}H^{(1)}_{3/2}(-k\eta)H^{(1)*}_{3/2}(-k\eta')\,,
\label{phiprop}
\end{equation}
where $H_{3/2}^{(1)}(-k\eta)$ with fixed $\nu=3/2$ appears due to the choice of the $d$-dimensional mass term~\eqref{eq:dDimensionalMassTerm}.
It is convenient to extract the momentum-conserving delta function from the correlator through the definition
\begin{equation}
    \cor{}{\phi(\eta_1,\vec k_1)...\phi(\eta_n,\vec k_n)}\equiv(2\pi)^{d-1}\delta^{(d-1)}\biggl(\sum_{i=1}^n\vec k_i\biggr)\cor{}{\phi(\eta_1,\vec k_1)...\phi(\eta_n,\vec k_n)}'\,.
    \label{corrconvention}
\end{equation}

For the following calculations, it is practical to represent the correlation functions diagrammatically.
However, as there are two different types of fields and vertices, corresponding to the forward and backward time branches, $+$ and $-$, as well as four different types of propagators, there exist numerous topologically identical diagrams featuring different ingredients.
Therefore, instead of introducing different symbols for all the vertices and propagators, and drawing all possible versions of the relevant diagrams corresponding to the in-in computation, only a single representative diagram for each family will be given.
This highlights the relevant diagram topology without burdening the discussion with the additional diagrams.
The scalar two-point function~\eqref{phiprop} is denoted by a solid line:
\begin{figure}[H]
    \centering
    \includegraphics[width=0.4\textwidth]{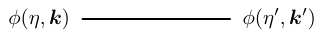}
\end{figure}
Finally, since in the Schwinger-Keldysh formalism, the two copies of the field appearing in the path integral are identified at the correlation time $\eta$, there is no need to specify whether the correlator contains fields living in the forward or backward time branch and one can simply write correlation functions of $\phi$.

\section{The method of regions}
\label{sec::regions}

The method of regions provides a strategy to perform the asymptotic expansion of 
Feynman integrals in a ratio of scales without evaluating the full integral. It was introduced in \cite{Beneke:1997zp} as a tool to compute the expansion 
of flat-space multi-loop Feynman integrals in non-relativistic kinematics. However, 
the method is quite general. It was soon extended to other kinematic limits 
\cite{Smirnov:1998vk} and has since found multiple applications.  
The method works as follows:
\begin{enumerate}[label=\arabic*)]
\item Identify all of the widely-separated external scales appearing in the 
integral. Then identify the relevant regions of the integral. A region 
is defined by the scaling of the integration variables with the small 
ratio(s) of external scales. Every integral contains a hard region, which 
corresponds to the naive expansion of the integrand in the small external 
quantities. However, the expansion of the {\em integral} is not the 
integral of the naive expansion of the integrand. The other regions are related 
to singularities of the {\em integrand}, i.e.~solutions to the Landau equations 
in the case of standard Feynman integrals, since it is the singularities 
that prevent the naive expansion.\footnote{When expressed in terms of the 
$\alpha$-representation, the regions of a Feynman integral can be 
given a geometric interpretation \cite{Pak:2010pt,Semenova:2018cwy,Gardi:2022khw}.}
\item Decompose the domain of integration of each integral into these regions, 
and Taylor-expand the integrand in  each region in all the quantities which are 
considered small with respect to others, accounting for the scaling of 
the integration variables.
\item Ignore the (fictitious) restriction on the integration boundaries 
introduced to decompose the integrals into regions and extend them to the entire 
integration domain of the full integral in every region. This guarantees that the contribution from 
each region is a series of terms with  homogeneous scaling with respect to the 
external scales and consequently each term contributes to a unique order 
in the asymptotic expansion after integration.  
\item Add up all regions and terms from the previous step. 
The result is the expansion of the original full 
integral. This is the essential point. While the notion of 
regions had already been employed in the formal justification of factorisation theorems \cite{Collins:1989gx}, the method introduced in 
\cite{Beneke:1997zp} turned it into a computational tool to obtain expansions 
of multi-scale integrals. 
\end{enumerate}
For the method to work, it is essential to introduce a dimensional or analytic 
regulator in all of the appearing integrals, even if they are finite. The reason for 
this is that the extension of the integration domains in step 3) is 
equivalent to adding scaleless integrals to the expression with 
restricted boundaries \cite{Beneke:1997zp,Jantzen:2011nz}, which vanish in 
analytic or dimensional regularisation, 
while the same is not true in other regularisation schemes. 
 
The individual integrals corresponding to the different regions of loop momenta are 
typically easier to evaluate than the full integral.
However, they are also usually more singular than the original expression and 
contain additional divergences which only cancel once all regions are summed. 
Sometimes it may be necessary to introduce an additional analytic regulator 
in the individual regions, even if the full integral is already regulated 
dimensionally \cite{Smirnov:1998vk}, as it happens for example 
with rapidity divergences in soft-collinear effective theory \cite{Beneke:2003pa}. 
Employing an analytical regulator is essential, as one still requires scaleless 
integrals to vanish.

The region expansion achieves the separation of the different scales appearing in the 
problem by re-expressing multi-scale integrals as sums over single-scale integrals. 
This procedure is closely related to the construction of an effective theory to 
describe the observables in a kinematic limit of interest, which 
is the reason why the method of regions has established itself as a powerful 
tool for matching computations in effective field theories.

In this paper, the method of regions is applied to the computation of late-time, 
in-in correlation functions of scalar fields in de Sitter space. 
The external momenta $k_i$ and the (conformal) correlation time $\eta$ play the roles of widely separated external scales through the requirement
\begin{equation}
    -k_i\eta\ll1\,,
\end{equation}
which defines the regions contributing to a given correlation function. 
As mentioned in the introduction, this simplifies the computations and provides 
some insight into the structure of the effective theory which describes the 
late-time limit of equal-time correlation functions in de Sitter 
space \cite{Cohen:2020php}. 

\section{Tree-level trispectrum in the full theory}
\label{treetrispec}
\begin{figure}
    \centering
    \includegraphics[width=0.52\textwidth]{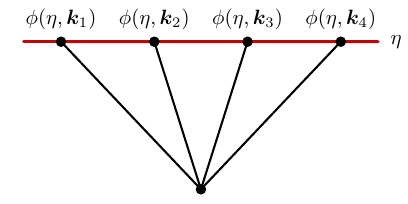}\qquad
    \includegraphics[width=0.4\textwidth]{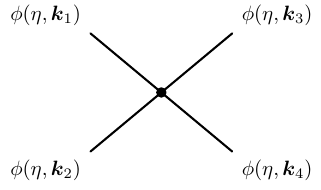}
    \caption{Diagrammatic representation of the tree-level trispectrum as a Witten diagram (left) and by its determining topology in the standard Feynman diagram depiction (right).
    The dark red line in the first diagram represents the late-time boundary where the correlator is evaluated.}
    \label{fig::4::WittenTopology}
\end{figure}

As a first demonstration of the method of regions consider the tree-level trispectrum
\begin{equation}
    \cor{}{\phi(\eta,\vec k_1)\phi(\eta,\vec k_2)\phi(\eta,\vec k_3)\phi(\eta,\vec k_4)}'\,.
\end{equation}
It corresponds to the Witten diagram as shown on the left-hand side in Fig.~\ref{fig::4::WittenTopology}. Alternatively, one can represent the above diagram by the basic diagram topology using a more standard depiction as seen on the right-hand side in the same figure.
The result for this late-time correlation function can be found e.g.~in \cite{Cohen:2021fzf} but is rederived here to serve as an example, highlighting the fact that already at tree-level, the in-in correlators have a non-trivial region structure.

%%%%%%%%%%%%%%%%%%%%%%%%%%%%%%%%%%%%%%%%%%%%%%%%%%%%%%%
%%%%%%%%%%%%%%%%%%%%%%%%%%%%%%%%%%%%%%%%%%%%%%%%%%%%%

\subsection{Direct computation}

In the in-in formalism, using the rules described in App.~\ref{SKrules}, the trispectrum reads
\begin{align}
&\cor{}{\phi(\eta,\vec k_1)\phi(\eta,\vec k_2)\phi(\eta,\vec k_3)\phi(\eta,\vec k_4)}\nonumber\\
&=i\tilde{\mu}^{4-d}\kappa\int_{-\infty}^{\eta}\frac{\der\eta'}{(-H\eta')^d}\;\int\biggl[\prod_{l=1}^4\frac{\der^{d-1}p_l}{(2\pi)^{d-1}}\biggr](2\pi)^{d-1}\delta^{(d-1)}\biggl(\sum_{l=1}^4\vec p_l\biggr)\nonumber\\
&\quad\times\biggl(\prod_{i=1}^4G_{--}(\eta,\vec k_i;\eta',\vec p_i)-\prod_{j=1}^4G_{++}(\eta,\vec k_j;\eta',\vec p_j)\biggr)\nonumber\\
&=i\tilde{\mu}^{4-d}\kappa(2\pi)^{d-1}\delta^{(d-1)}\biggl(\sum_{i=1}^4\vec k_i\biggr)\nonumber\\
&\quad\times\int_{-\infty}^{\eta}\frac{\der\eta'}{(-H\eta')^d}\;\biggl(\prod_{i=1}^4\cor{}{\phi(\eta',\vec k_i)\phi(\eta,-\vec k_i)}'
-\prod_{j=1}^4\cor{}{\phi(\eta,\vec k_j)\phi(\eta',-\vec k_j)}'\biggr)\,.
\end{align}
The second equality follows from the definitions of the propagators~\eqref{ining}, using the delta-functions contained in the definition of the Wightman function~\eqref{phiprop} to eliminate the momentum integrals and introducing the primed correlator~\eqref{corrconvention}.
The remaining two-point functions can be expressed in terms of the $d$-dimensional mode functions~\eqref{eq::ddimensionalModeFunction}, which contain the $\nu=3/2$ Hankel functions~\eqref{eq:MasslessHankelFunction} due to the $d$-dimensional mass term~\eqref{eq:dDimensionalMassTerm}.
This results in
\begin{align}
&\cor{}{\phi(\eta,\vec k_1)\phi(\eta,\vec k_2)\phi(\eta,\vec k_3)\phi(\eta,\vec k_4)}'\nonumber\\
&=\frac{\tilde{\mu}^{4-d}\kappa H^d(-H\eta)^{2d-8}}{8(k_1k_2k_3k_4)^3}\Im\biggl[\int_{-\infty}^{\eta}\frac{\der\eta'}{(-\eta')^{8-d}}e^{ik_t(\eta'-\eta)}\prod_{j=1}^4(-i+k_j\eta)(i+k_j\eta')\biggr]\,.
\label{triint}
\end{align}
Since the appearing time integrals are finite, the limit $d\rightarrow4$ commutes with the integration and one can evaluate the integrals directly after setting $d=4$. 
They can be computed exactly in terms of incomplete gamma functions.
Taking the late-time limit $-k_i\eta\rightarrow0$ results in~\cite{Cohen:2021fzf}
\begin{align}
&\lim\limits_{-k_i\eta\rightarrow0}\cor{}{\phi(\eta,\vec k_1)\phi(\eta,\vec k_2)\phi(\eta,\vec k_3)\phi(\eta,\vec k_4)}'\nonumber\\
&=\frac{\kappa H^4}{8(k_1k_2k_3k_4)^3}\biggl[\frac{1}{3}\biggl(\frac{1}{3}+\ln(-e^{\gamma_E}k_t\eta)\biggr)\sum_{j=1}^4k^3_j-\frac{k_1k_2k_3k_4}{k_t}+k_t\biggl(\sum_{j<l}k_jk_l-\frac{4}{9}k^2_t\biggr)\biggr]\,,
\label{fulltri}
\end{align}
with $k_t\equiv k_1+k_2+k_3+k_4$ the sum of 
the moduli of the external momenta.

\subsection{Late-time correlator via the method of regions}
\label{4ptregions}

To apply the method of regions to this correlator, one first needs to identify the relevant regions of the time integral.
Then, the integral can be decomposed into simpler single-scale integrals, and the result~\eqref{fulltri} arises as the sum of all regions.

A peculiar feature of in-in correlation functions is that they present a non-trivial region structure already at tree level, due to the time integral that is present for each vertex. 
From this point of view, late-time, in-in correlation functions at tree level resemble momentum-space loop-level correlators in flat space. 
In the language of SdSET, this translates to the matching of correlation functions at tree level, which requires the specification of ``non-Gaussian initial conditions"~\cite{Cohen:2020php}.
At loop level, the interplay of the different regions of time and momentum integrals results in a richer structure than might have been naively expected, which then carries over to SdSET. The discussion of the simultaneous decomposition of time and momentum integrals is, however, postponed to Sec.~\ref{sec::OneLoopMixing}.

\subsubsection{Identification of the regions}
\label{triregions}
To determine the relevant regions of the time integral appearing in~\eqref{triint}, consider the general integral
\begin{equation}
    \int_{-\infty}^{\eta}\der\eta'\;(-\eta')^{-\ve}f(-k_i\eta,-k_i\eta')\,,
    \label{morexample}
\end{equation}
understood to be evaluated in the late-time limit $-k_i\eta\to0$.
The factor $(-\eta')^{-\ve}$ represents the effect of the dimensional regulator while the function $f$ is the remainder of the integrand, which in this case depends only on the combinations $-k_i\eta$ or $-k_i\eta^\prime$.
One can now employ the method of regions following Sec.~\ref{sec::regions}.

In the first step, one determines the relevant external scales of the problem.
In this example, there is the combination $-k_i\eta$, which is assumed to be small, $-k_i\eta\ll1$.
This implies the widely-separated scales $\lvert\eta\rvert\ll k_i^{-1}$, which in turn define the regions of the time integral in~\eqref{morexample} through the typical scaling of $\eta^\prime$.
First, there is the late-time region where $\eta^\prime\sim\eta$, and the integrand can be expanded in both the external small scale $-k_i\eta\ll1$ and the combination $-k_i\eta^\prime\ll1$ containing the integration variable.
Second, there is the early-time region $\eta^\prime\sim -k_i^{-1}\ll\eta$.
While the integrand can be expanded in the external scale $-k_i\eta\ll1$, one has to treat $-k_i\eta^\prime\sim1$ and cannot perform an expansion in this argument.

In the next step, one decomposes the time integral~\eqref{morexample} into the two regions.
Since the upper integration boundary is fixed to $\eta$, one can extend the integral to the full domain $\eta^\prime\in(-\infty,0)$, corresponding to integration from past to future infinity, through a $\theta$-function as
\begin{equation}\label{eq::4::ExampleWithTheta}
    \int_{-\infty}^{\eta}\der\eta'\;(-\eta')^{-\ve}f(-k_i\eta,-k_i\eta') = \int_{-\infty}^0 \der \eta^\prime\:\theta(\eta-\eta^\prime)(-\eta')^{-\ve}f(-k_i\eta,-k_i\eta')\,.
\end{equation}
This integration is now split into both regions by introducing the intermediate scale $\Lambda$ as $-k^{-1}_i\ll\Lambda\ll\eta$,
so that the integral~\eqref{eq::4::ExampleWithTheta} reads
\begin{align}
    &\int_{-\infty}^{0}\der\eta'\;\theta(\eta-\eta^\prime)(-\eta')^{-\ve}f(-k_i\eta,-k_i\eta')\nonumber\\
    &=\int_{-\infty}^{\Lambda}\der\eta'\;\theta(\eta-\eta^\prime)(-\eta')^{-\ve}f(-k_i\eta,-k_i\eta')+\int_{\Lambda}^{0}\der\eta'\;\theta(\eta-\eta^\prime)(-\eta')^{-\ve}f(-k_i\eta,-k_i\eta')\nonumber\\
    &\equiv I_{\mathrm{early}} + I_{\mathrm{late}}\,.
    \end{align}
The first integral in the second line corresponds to the early-time region $\eta^\prime~\sim -k_i^{-1}\ll \eta$. Consequently, one can Taylor-expand as $\eta^\prime\ll\eta$ and $-k_i\eta\ll1$ but has to keep the full dependence on $-k_i\eta^\prime$.
The $\theta$ function is expanded in a distributional sense as
\begin{equation}
    \theta(\eta-\eta^\prime) = \theta(-\eta^\prime) + \eta\, \delta(-\eta^\prime) + \mathcal{O}(\eta^2)\,, 
\end{equation}
where only the first term is relevant to obtain the leading-order terms of the asymptotic expansion as $\eta\to 0$.
One can therefore rewrite the early-time integral as
\begin{equation}
    I_{\mathrm{early}} = \int_{-\infty}^\Lambda\der\eta'\;\theta(-\eta^\prime)(-\eta')^{-\ve}f(-k_i\eta,-k_i\eta')\Bigr|_{\substack{-k_i\eta'\sim 1\\-k_i\eta\ll1}}\,.
\end{equation}
The second integral corresponds to the late-time region where $\eta^\prime\sim\eta$. This implies that one has to keep $\theta(\eta-\eta^\prime)$, as the argument is homogeneous, but one can perform the Taylor-expansion in the small quantities $-k_i\eta$ and $-k_i\eta^\prime$.
Consequently, this integral reads
\begin{equation}
    I_{\mathrm{late}} = \int_{\Lambda}^{0}\der\eta'\;\theta(\eta-\eta^\prime)(-\eta')^{-\ve}f(-k_i\eta,-k_i\eta')\Bigr|_{\substack{-k_i\eta'\ll 1\\-k_i\eta\ll1}}\,.
\end{equation}

In step 3, one sets $\Lambda\to0$ ($\Lambda\to -\infty$) for the early-time (late-time) integral, thereby removing the intermediate scale. This amounts to the addition of scaleless integrals which vanish in dimensional regularisation. One can now also include the restrictions imposed by the $\theta$-function, which have a non-trivial effect on the domain of integration only in the late-time integral, resulting in
\begin{align}\label{eq::4::ExampleRegionsFinal}
    &\int_{-\infty}^{0}\der\eta'\;\theta(\eta-\eta^\prime)(-\eta')^{-\ve}f(-k_i\eta,-k_i\eta')\\
    &=\int_{-\infty}^0\der\eta'\;(-\eta')^{-\ve}f(-k_i\eta,-k_i\eta')\Bigr|_{\substack{-k_i\eta'\sim 1\\-k_i\eta\ll1}}+\int_{-\infty}^{\eta}\der\eta'\;(-\eta')^{-\ve}f(-k_i\eta,-k_i\eta')\Bigr|_{\substack{-k_i\eta'\ll 1\\-k_i\eta\ll1}}\,.\nonumber
\end{align}
Due to the Taylor expansion, each integral now contains a (simpler) integrand where each term scales homogeneously with respect to the external scales.
However, as anticipated, both integrals are now divergent, whereas the original integral~\eqref{morexample} was finite.
The first term, the early-time integral, is divergent as $\eta^\prime\to0$ due to the expansion $\eta^\prime\ll\eta$, thus generating an IR pole in $\varepsilon$.
Moreover, since the integration is now over the full domain and the integrand is expanded in $-k_i\eta$, the integral yields a time-independent contribution.

The second integral, on the other hand, is divergent for $\eta^\prime\to-\infty$ due to the Taylor expansion for $-k_i\eta^\prime\ll 1$, resulting in an ultraviolet (UV) pole in $\varepsilon$.
Both divergences cancel in the sum, as explicitly verified for this example below.
However, before this cancellation, the poles in $\ve$ ensure that the $\ve$-expansion of the two summands generates the large late-time logarithm  in~\eqref{fulltri}.

\subsubsection{Late-time region}

In the late-time region, both $-k_i\eta$ and $-k_i\eta^\prime$ are treated as small parameters.
The product in the integrand~\eqref{triint} must be expanded up to $\Lo((-k_i\eta')^{3})$ and $\Lo((-k_i\eta)^3)$ to obtain the leading (non-vanishing) contribution of the asymptotic expansion.
One finds
\begin{align}
&\cor{}{\phi(\eta,\vec k_1)\phi(\eta,\vec k_2)\phi(\eta,\vec k_3)\phi(\eta,\vec k_4)}'\Bigr|_{\textrm{late }\eta'}\nonumber\\
&=\frac{\tilde{\mu}^{4-d}\kappa H^d(-H\eta)^{2d-8}}{8(k_1k_2k_3k_4)^3}\lim\limits_{\eta\rightarrow0}\Im\biggl[\biggl(1+\frac{\eta^2}{2}\sum_{i=1}^4k^2_i-\frac{i\eta^3}{3}\sum_{i=1}^4k^3_i + \mathcal{O}\Bigl((-k_i\eta)^4\bigr)\biggr)\nonumber\\
&\quad\times\int_{-\infty}^{\eta}\frac{\der\eta'}{(-\eta')^{8-d}}\biggl(1+\frac{(-\eta')^2}{2}\sum_{i=1}^4k^2_i-\frac{i(-\eta')^3}{3}\sum_{i=1}^4k^3_i + \mathcal{O}\Bigl((-k_i\eta^\prime)^4\bigr)\biggr)\biggr]\,.
\label{softtriexp}
\end{align}
All appearing integrals are of the simple form
\begin{equation}
\int_{-\infty}^{\eta}\frac{\der\eta'}{(-\eta')^a}=\frac{(-\eta)^{1-a}}{a-1}\,,
\end{equation}
which scale homogeneously in $\eta$. 
In particular, one can already determine the complete non-analytic dependence on the external scales of the integral before doing any computations.
Since the integrand is Taylor-expanded in both $-k_i\eta$ and $-k_i\eta^\prime$, there can be no non-analytic dependence on the external momenta $k_i$, and the only such dependence that remains must stem from the factor $(-\eta^\prime)^{-2\varepsilon}$.
Therefore, the result of the integral must be proportional to $(-\eta)^{-2\ve}$.
After expanding this result in $\ve$, this factor gives rise to the time-dependent part of the logarithm in~\eqref{fulltri}.

The only terms that contribute to~\eqref{softtriexp} in the limit $\eta\rightarrow0$ are the ones proportional to $\eta^3(-\eta')^{d-8}$ and $(-\eta')^{d-5}$.
The integral is UV-divergent, since the latter term diverges logarithmically as $\eta\to-\infty$. 
This explains why the result must be proportional to the sum over cubes of $k_i$.  
Evaluating the integrals, one obtains to leading order in $-k_i\eta$
\begin{align}
&\cor{}{\phi(\eta,\vec k_1)\phi(\eta,\vec k_2)\phi(\eta,\vec k_3)\phi(\eta,\vec k_4)}'\Bigr|_{\textrm{late }\eta'}\nonumber\\
&=-\frac{\kappa H^{4+2\ve}(-H\eta)^{-4\ve}}{8(k_1k_2k_3k_4)^3}\biggl(-\frac{H^2\eta}{\tilde{\mu}}\biggr)^{\!-2\ve}\frac{1}{2\ve(3+2\ve)}\sum_{j=1}^4k^3_j\nonumber\\
&=\frac{\kappa H^{4+2\ve}(-H\eta)^{-4\ve}}{24(k_1k_2k_3k_4)^3}\,\biggl[-\frac{1}{2\ve}+\frac{1}{3}+\ln\biggl(-\frac{H^2\eta}{\tilde\mu}\biggr)\biggr]\sum_{j=1}^4k^3_j + \mathcal{O}(\varepsilon)\,.
\label{softtri}
\end{align} 
It is convenient to leave the overall $\ve$-dependent prefactor unexpanded since it appears for both regions and thus drops out in the sum at the end when the poles cancel.\footnote{
The $\ve$-dependent prefactor arises as follows: $H^{4+2\ve}$ corresponds to the overall dimension of the correlation function in $d$ dimensions, and the factor $(-H\eta)^{-4\ve}$ is associated with the external mode functions. The $(-\eta)^{-2\ve}$ from the region integral then combines with $\tilde{\mu}^{2\ve}$ and the remaining powers of $H$ to the dimensionless ratio $-H^2\eta/\tilde{\mu}$.
}
The argument of the logarithm in~\eqref{softtri} contains the ratio $\tilde\mu/\eta$.
This is to be expected, since the recasting of the original integral into simpler, single-scale ones factorises the logarithms of widely separated scales, and $\tilde\mu$ takes the role of the factorisation scale.
This logarithm can be expressed as
\begin{equation}
    \ln\biggl(-\frac{H^2\eta}{\tilde\mu}\biggr) = \ln\biggl(\frac{H}{a(\eta)\tilde\mu}\biggr)\,,
\end{equation}
which is small for $\tilde\mu\sim -H^2\eta = H/a(\eta)$.
This scale choice is consistent with the natural scale of the late-time region, where the superhorizon modes satisfy $k_{i,\mathrm{phys}}\ll H$.
Moreover, one can use the dependence on $\tilde\mu$ to track the $\eta$-dependence of the late-time limit of the full correlation function.
This offers the possibility of controlling the secular logarithms.

\subsubsection{Early-time region}
In the early-time region, one expands the integrand of~\eqref{triint} in the small quantity $-k_i\eta\ll1$ but keeps the full dependence on $-k_i\eta^\prime$. One also sets $\eta\to0$ in the integral boundary following~\eqref{eq::4::ExampleRegionsFinal}.
This results in
\begin{align}
&\cor{}{\phi(\eta,\vec k_1)\phi(\eta,\vec k_2)\phi(\eta,\vec k_3)\phi(\eta,\vec k_4)}'\Bigr|_{\textrm{early }\eta'}\nonumber\\
&=\frac{\tilde{\mu}^{4-d}\kappa H^d(-H\eta)^{2d-8}}{8(k_1k_2k_3k_4)^3}\Im\biggl[\int_{-\infty}^0\frac{\der\eta'}{(-\eta')^{8-d}}e^{ik_t\eta'}\prod_{j=1}^4(i+k_j\eta')\biggr]\,.\label{momprod}
\end{align}
As expected, this expression is time-independent and the remaining integrals are single-scale integrals of the form
\begin{equation}
    \int_{-\infty}^0\frac{\der\eta'}{(-\eta')^a}\;e^{ik_t\eta'}=(ik_t)^{a-1}\Gamma(1-a)\,,
\end{equation}
which scale homogeneously in the only remaining external scale $k_t$. The only non-analytic dependence generated by them is $k_t^{2\ve}$, which explains the appearance of $k_t$ inside the logarithm of~\eqref{fulltri}.
After evaluating the integrals in~\eqref{momprod}, one finds
\begin{align}
&\cor{}{\phi(\eta,\vec k_1)\phi(\eta,\vec k_2)\phi(\eta,\vec k_3)\phi(\eta,\vec k_4)}'\Bigr|_{\textrm{early }\eta'}\nonumber\\
&=\frac{\kappa H^{4+2\ve}(-H\eta)^{-4\ve}}{8(k_1k_2k_3k_4)^3}\Gamma(-3-2\ve)\biggl(\frac{k_t\tilde{\mu}}{H^2}\biggr)^{\!2\ve}\biggl\{2\sum_{j=1}^4k^3_j+\ve\biggl[\frac{22}{3}\sum_{j=1}^4k^3_j\nonumber\\
&\hspace{0.4cm}-12\biggl(\frac{k_1k_2k_3k_4}{k_t}-k_t\biggl(\sum_{j<l}k_jk_l-\frac{4}{9}k^2_t\biggr)\biggr)\biggr]+\Lo(\ve^2)\biggr\}\,.
\end{align}
%where terms that vanish in the limit $\ve\rightarrow0$ were dropped.
The $\Gamma(-3-2\ve)$ contains the infrared pole from $\eta\to0$.
As anticipated, the overall non-analytic dependence is $k_t^{2\varepsilon}$, while the remaining combinations of momenta appearing in the bracket are generated by the integer powers of $k_t$ from the time integrals, together with the momenta already present in the product in~\eqref{momprod}. 
The $\ve$-expansion reads
\begin{align}
&\cor{}{\phi(\eta,\vec k_1)\phi(\eta,\vec k_2)\phi(\eta,\vec k_3)\phi(\eta,\vec k_4)}'\Bigr|_{\textrm{early }\eta'}\nonumber\\
&=\frac{\kappa H^{4+2\ve}(-\eta H)^{-4\ve}}{8(k_1k_2k_3k_4)^3}
\biggl\{\frac{1}{3}\biggl[\frac{1}{2\ve}+\ln\biggl(\frac{e^{\gamma_E}k_t\tilde\mu}{H^2}\biggr)\biggr]\sum_{j=1}^4k^3_j-\frac{k_1k_2k_3k_4}{k_t}\nonumber\\
&\quad+k_t\biggl(\sum_{j<l}k_jk_l-\frac{4}{9}k^2_t\biggr)\biggr\}\,.
\label{hardtri}
\end{align}
In this region, the logarithm has the form $\ln((k_t\tilde\mu)/H^2)$.
After identifying $k_t\sim -\eta_{\mathrm{cr},k_t}^{-1}$, the time when $k_t$ crosses the horizon, one notices that this is the same logarithm as in~\eqref{softtri}, now with $\eta\to\eta_{\mathrm{cr},k_t}$.
This logarithm is small for $\tilde\mu\sim -H^2\eta_{\mathrm{cr},k_t}=H/a(\eta_{\mathrm{cr},k_t})$.

\subsubsection{Sum of the regions}
The sum of both regions yields
\begin{align}
&\cor{}{\phi(\eta,\vec k_1)\phi(\eta,\vec k_2)\phi(\eta,\vec k_3)\phi(\eta,\vec k_4)}'\Bigr|_{\textrm{early }\eta'}+\cor{}{\phi(\eta,\vec k_1)\phi(\eta,\vec k_2)\phi(\eta,\vec k_3)\phi(\eta,\vec k_4)}'\Bigr|_{\textrm{late }\eta'}\nonumber\\
&=\frac{\kappa H^4}{8(k_1k_2k_3k_4)^3}\biggl[\frac{1}{3}\biggl(\frac{1}{3}+\ln(-e^{\gamma_E}k_t\eta)\biggr)\sum_{j=1}^4k^3_j-\frac{k_1k_2k_3k_4}{k_t}+k_t\biggl(\sum_{j<l}k_jk_l-\frac{4}{9}k^2_t\biggr)\biggr]\,.
\end{align}
As expected, the poles in $\ve$ cancel and, consequently, the global prefactor that was not expanded simply goes to 1 as $\ve\to 0$.
Moreover, this result directly reproduces the late-time limit of the tree-level trispectrum~\eqref{fulltri}.

While the result can simply be obtained from a direct computation of the full integral, this exercise serves as an instructive example highlighting the important features of the method of regions in de Sitter space.
The advantage of the region expansion becomes evident when considering less trivial diagrams, in particular loop diagrams, where the function-space of the full calculation is more complicated, as for the correlator discussed in Sec.~\ref{sec::OneLoopMixing}.

\section{Late-time, one-loop power spectrum}
\label{powerspec}

As the next example, consider the one-loop correction to the power spectrum
\begin{equation}
    \cor{}{\phi(\eta,\vec k)\phi(\eta,-\vec k)}'
\end{equation}
in the late-time limit. 
At $\Lo(\kappa)$, there are two contributions, one from the loop diagrams and one from the mass counterterm $\delta m^2$, which is required to remove the UV divergence of the loop integral. Both topologies are depicted in Fig.~\ref{fig::PowerSpectrumOneLoop}.
Like in flat space, there is no contribution from the field renormalisation and $Z_\phi= 1 + \mathcal{O}(\kappa^2)$.
The finite part of $\delta m^2$ can only be determined after choosing a suitable renormalisation condition for the mass term, which is discussed in detail below. 
It is well-known that the one-loop power spectrum of the massless scalar field contains also an IR-divergence, which is regulated by dimensional regularisation in the present approach.
To the best of our knowledge, this computation has not been performed in this scheme in the literature.
As in the previous section, we first present the direct computation of the one-loop correction and then reproduce the result via the method of regions.

\subsection{Direct computation}
\label{full1loop}

\subsubsection{One-loop diagram}

\begin{figure}
    \centering
    \includegraphics[width=0.49\textwidth]{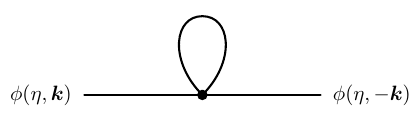}
    \quad
    \includegraphics[width=0.42\textwidth]{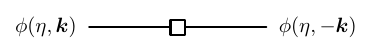}
    \caption{Topology of Feynman diagrams contributing to the power-spectrum at the one-loop level due to the $\kappa\phi^4$-interaction (left) and the mass counterterm (right).}
    \label{fig::PowerSpectrumOneLoop}
\end{figure}

The one-loop correction to the power spectrum corresponds to a single tadpole diagram, as shown on the left in Fig.~\ref{fig::PowerSpectrumOneLoop}.
It is given by
\begin{align}\label{eq:PowerSpectrumStart}
&\cor{}{\phi(\eta,\vec k)\phi(\eta,-\vec k)}'\Bigr|_{\Lo(\kappa)}
=\frac{\tilde{\mu}^{4-d}\kappa H^{d-2}}{8}(-H\eta)^{d-4}\nonumber\\
&\hspace*{1cm}\times\,\Im\biggl[\int_{-\infty}^{\eta}\frac{\der\eta'}{(-\eta')^{8-d}}\frac{e^{2ik(\eta'-\eta)}(i-k\eta)^2(i+k\eta')^2}{k^6}\biggr]\int\frac{\der^{d-1}\vec{l}}{(2\pi)^{d-1}}\frac{1+l^2{\eta'}^2}{l^3}\,.
\end{align}
This contribution has a simple form, as the time and momentum integrals factorise.
In addition, the time integral is finite, while the
momentum integral is both UV and IR divergent and scaleless.
The one-loop correction to the power spectrum therefore vanishes in dimensional regularisation.

However, its UV divergence is required to determine the mass counterterm $\delta m^2$.
As this divergence is independent of the details of the IR physics, one can choose any IR regulator to extract this piece.
For the following computation, we isolate the UV pole by introducing a (comoving) IR cutoff $\Tilde{\Lambda}$ through the replacement
\begin{equation}
\frac{1}{l^3}\rightarrow  \frac{1}{[l^2+\tilde{\Lambda}^2]^{\frac{3}{2}}}=\frac{1}{[l^2+a(\eta)^2\Lambda^2]^{\frac{3}{2}}}\,.
\label{IRmod}
\end{equation}
To obtain the second equality, we have chosen a corresponding IR-cutoff $\Lambda$ for the physical (as opposed to comoving) loop momentum at correlation time $\eta$.
The introduction of $\Lambda$ inevitably leads to a result which breaks de Sitter-invariance. 
However, the UV-pole of the integral is independent of the regulator.

After introducing the IR cutoff, the correlator~\eqref{eq:PowerSpectrumStart} reads
\begin{align}
&\cor{}{\phi(\eta,\vec k)\phi(\eta,-\vec k)}'\Bigr|_{\Lo(\kappa)}=\frac{\tilde{\mu}^{4-d}\kappa H^{d-2}}{8}(-H\eta)^{d-4}\nonumber\\
&\hspace*{0.5cm}\times\,\Im\biggl[\int_{-\infty}^{\eta}\frac{\der\eta'}{(-\eta')^{8-d}}\frac{e^{2ik(\eta'-\eta)}(i-k\eta)^2(i+k\eta')^2}{k^6}\biggr]
\,\int\frac{\der^{d-1}\vec{l}}{(2\pi)^{d-1}}\frac{1+l^2{\eta'}^2}{[l^2+a(\eta)^2\Lambda^2]^{\frac{3}{2}}}\,.
\label{modexpr}
\end{align}
One can immediately drop the second summand in the $\vec l$-integral, as it yields a term proportional to $\Lambda^2$, which vanishes as $\Lambda\rightarrow0$. 
After performing the $\eta'$- and $\vec l$-integrations and taking the limit $-k\eta\rightarrow0$, one obtains
\begin{align}
&\lim\limits_{-k\eta\rightarrow0}\cor{}{\phi(\eta,\vec k)\phi(\eta,-\vec k)}'\Bigr|_{\Lo(\kappa)}\nonumber\\
&=\frac{\kappa H^2}{24\pi^2k^3}(-H\eta)^{-2\ve}\biggl(\frac{e^{\gamma_E}\mu}{\Lambda}\biggr)^{\!2\ve}\biggl[-\frac{1}{2\ve}\Bigl[2-\ln(-2e^{\gamma_E}k\eta)\Bigr]+\frac{1}{2}\ln^2(-2e^{\gamma_E}k\eta)\nonumber\\
&\quad-\frac{7}{3}\ln(-2e^{\gamma_E}k\eta)+\frac{8}{3}-\frac{\pi^2}{24}\biggr]\,.
\label{eq::oneloop2ptreg}
\end{align}
This result contains a single pole in $\ve$, which corresponds to the dimensionally-regulated UV-divergence of the momentum integral.
It must be cancelled through the mass counterterm $\delta m^2$. 
The IR-pole of the integral instead appears as a logarithmic divergence in the limit $\Lambda\rightarrow0$ due to the modification~\eqref{IRmod}.

\subsubsection{Mass-counterterm diagram}
\label{sec::mcounter}

The contribution from the mass counterterm reads
\begin{equation}\label{eq:5:MassCountertermFull}
\cor{}{\phi(\eta,\vec k)\phi(\eta,-\vec k)}'\Bigr|_{\delta m^2}=\frac{\delta m^2}{2}(-H\eta)^{d-4}\Im\biggl[\int_{-\infty}^{\eta}\frac{\der \eta'}{(-\eta')^{4}}\frac{e^{2ik(\eta'-\eta)}(i-k\eta)^2(i+k\eta')^2}{k^6}\biggr]\,.
\end{equation}
Curiously, in dimensional regularisation, the $\varepsilon$-dependence of the propagator and the measure of the action conspire so that the time integral is the same in all dimensions, i.e.~is independent of $\varepsilon$. 
Since the integral is finite, this is not a problem for the present computation. 
However, when decomposing it into regions, this necessitates the introduction of an additional analytic regulator, as is discussed below.\footnote{This is an accidental cancellation which occurs in the present regularisation scheme. It can easily be checked that in other schemes, such as analytic regularisation, this is not the case.}
After evaluating the time integral and expanding the result in the late-time limit, one obtains
\begin{equation}
\lim\limits_{-k\eta\rightarrow0}\cor{}{\phi(\eta,\vec k)\phi(\eta,-\vec k)}'\Bigr|_{\delta m^2}=-\frac{\delta m^2}{3k^3}(-H\eta)^{-2\ve}\Bigl[2-\ln(-2e^{\gamma_E}k\eta)\Bigr]\,.
\label{eq::dmintermediate}
\end{equation}
To remove the UV divergence of~\eqref{eq::oneloop2ptreg}, one can choose
\begin{equation}
\delta m^2=\frac{\kappa H^2}{8\pi^2}\biggl[-\frac{1}{2\ve}+\delta \hat m^2_{\textrm{fin.}}\biggr]\,,
\label{eq::dm}
\end{equation}
with an undetermined finite part $\delta\hat m^2_{\textrm{fin}}$ 
whose precise form depends on the specific renormalisation condition for the mass term.

\subsubsection{Renormalised one-loop result, finite part of the counterterm}

As the mass counterterm is now determined and the UV pole is cancelled, one can let $\Lambda\rightarrow0$ in~\eqref{modexpr}. In this case, the regularised one-loop two-point function vanishes, and the renormalised one-loop correction to the power spectrum at late times is given by
\begin{align}
&\lim\limits_{-k\eta\rightarrow0}\biggl[\cor{}{\phi(\eta,\vec k)\phi(\eta,-\vec k)}'\Bigr|_{\Lo(\kappa)}+\cor{}{\phi(\eta,\vec k)\phi(\eta,-\vec k)}'\Bigr|_{\delta m^2}\biggr]\nonumber\\
&=\frac{\kappa H^2(-H\eta)^{-2\ve}}{24\pi^2k^3}\Bigl[2-\ln(-2ke^{\gamma_E}\eta)\Bigr]\biggl[\frac{1}{2\ve}-\delta\hat m^2_{\textrm{fin}}\biggr]\,,
\label{eq::pwrren}
\end{align}
i.e.~essentially by the mass-counterterm contribution.
However, the left-over pole should now be viewed as the IR-pole of $\cor{}{\phi(\eta,\vec k)\phi(\eta,-\vec k)}'|_{\Lo(\kappa)}$, since the UV-pole is cancelled by the mass-counterterm by construction.

The above result also contains the undetermined finite constant $\delta\hat m^2_{\textrm{fin}}$. To completely fix this finite piece, one has to specify a renormalisation condition for the mass term. In flat-space 
$\phi^4$-theory, the masslessness of the scalar field is naturally associated with the choice of $\delta m^2$ that keeps the pole of the time-ordered two-point function at $p^2=0$. In de~Sitter space, the analogue of the on-shell mass renormalisation condition would be that the power spectrum remains unaltered by loop corrections at a particular value of $k$ in the IR, e.g.~$k\to 0$. Clearly, this condition cannot be imposed perturbatively, as this inevitably leads to an infrared-divergent contribution to $\delta\hat m^2_{\textrm{fin}}$ which contains the explicit IR pole.
However, the IR-divergences that appear in perturbative computations are just a symptom of the fact that there is no de~Sitter-invariant vacuum state for the \emph{free} massless, minimally coupled scalar field~\cite{Allen:1985ux}, so the usual perturbative expansion is not well-defined.
In turn, the correlation functions defined in this perturbation theory do not qualify as sensible physical observables, and should not be used to formulate renormalisation conditions for the theory.
Instead, the conditions should be imposed on the non-perturbative (``resummed'') correlation functions, which are IR-finite.
For the specific case of the mass renormalisation, one such criterion might be that the resummed power-spectrum remains unaltered relative to the leading-order resummed value in the limit $k\to0$.
This should result in an IR-finite 
$\delta\hat m^2_{\textrm{fin}}$, which constitutes the analogue of flat-space on-shell mass renormalisation.

One example of such a non-perturbative framework for the computation of correlation functions is the extension of the stochastic formalism to SdSET~\cite{Cohen:2021fzf}. 
In this context, $\delta\hat m^2_{\textrm{fin}}$ is incorporated into one of the Wilson coefficients of the EFT by the matching procedure. 
These coefficients are in turn related to the coefficients of the Kramers-Moyal equation, which acts as an effective equation of motion for correlation functions in the EFT, generalising the Fokker-Planck equation of the stochastic formulation. 
Correlators computed using the solution of the Kramers-Moyal equation are IR-finite and well-behaved in the limit $\eta\rightarrow0$. 
They retain a dependence on $\delta\hat m^2_{\textrm{fin}}$ and are thus suitable objects on which to impose a renormalisation condition to determine this finite piece.\footnote{
In the IR limit, the UV contribution to the self-mass squared is subleading relative to the dynamically generated mass by a power of $\sqrt{\kappa}$, see discussion of Figure~1 in \cite{Beneke:2012kn}.}
In the following, $\delta\hat m^2_{\textrm{fin}}$ is therefore simply kept as a variable, which could be fixed by the proper procedure described above.

\subsection{One-loop power spectrum via the method of regions}
\label{sec:OneLoopRegions}

One can apply the method of regions to reproduce the late-time results~\eqref{eq::oneloop2ptreg} and~\eqref{eq::dmintermediate}. 
Compared to the tree-level four-point function in Sec.~\ref{4ptregions}, the correlator now contains two integrals, as there is the additional momentum integral.
However, this integral is scaleless, so it does not contribute any regions of its own. The calculation is thus analogous to the tree-level trispectrum considered above.
For the application of the method of regions, this section does not provide any new insights.
However, the results derived here are essential for the matching onto SdSET.
Therefore, the explicit computations are performed in App.~\ref{app:PowerSpectrum}, and only a quick summary is presented here.

In the first step, we determine the relevant regions contributing to the regularised one-loop integral~\eqref{modexpr}.
For the time integral, in the late-time limit $-k\eta\to 0$, the external scales for the time integral are determined by $-k\eta\ll1$, and one again finds two regions, as was the case for the four-point function in Sec.~\ref{4ptregions}.
The additional momentum integral is scaleless and vanishes, so it does not contribute additional regions.
However, in the full theory, the UV divergence of the momentum integral is required to determine the counterterm $\delta m^2$.
Likewise, the UV divergence of the late-time, soft-momentum region is important in the renormalisation of SdSET.
Therefore, we make use of the same IR cutoff $\Lambda$ as in the direct computation~\eqref{IRmod} and define the regulated momentum integral as
\begin{equation}
    I[a(\eta)\Lambda]\equiv\int\frac{\der^{d-1}l}{(2\pi)^{d-1}}\frac{1}{[l^2+a(\eta)^2\Lambda^2]^{\frac{3}{2}}}
\end{equation}
to write the one-loop contribution as
\begin{align}
&\cor{}{\phi(\eta,\vec k)\phi(\eta,-\vec k)}'\Bigr|_{\Lo(\kappa)}=\frac{\tilde{\mu}^{4-d}\kappa H^{d-2}}{8}(-H\eta)^{d-4}\nonumber\\
&\hspace*{0.5cm}\times\,\Im\biggl[\int_{-\infty}^{\eta}\frac{\der\eta'}{(-\eta')^{8-d}}\frac{e^{2ik(\eta'-\eta)}(i-k\eta)^2(i+k\eta')^2}{k^6}\biggr]\, I[a(\eta)\Lambda]\,.
\label{1looptint}
\end{align}
The same region analysis also applies to the mass counterterm~\eqref{eq:5:MassCountertermFull}, which features only a time integral.
As previously noted, in dimensional regularisation, the $d$-dependence drops out of the integral. 
Therefore, to separate the time integral into regions, we introduce the additional analytic regulator $(-\nu\eta')^{-2\delta}$ in the time integral, where $\nu$ is a (comoving) scale with mass dimension 1.
This regulator explicitly breaks de~Sitter-invariance. However, as the expansion in $\delta$ is performed before the $\ve$-expansion, and the poles in $\delta$ must cancel between both regions, the invariance is restored in the final sum of both regions.
The counterterm contribution then reads
\begin{equation}
\cor{}{\phi(\eta,\vec k)\phi(\eta,-\vec k)}'\Bigr|_{\delta m^2}\!\!\!=\frac{\delta m^2}{2}(-H\eta)^{d-4}\nu^{-2\delta}\Im\biggl[\int_{-\infty}^{\eta}\frac{\der \eta'}{(-\eta')^{4+2\delta}}\frac{e^{2ik(\eta'-\eta)}(i-k\eta)^2(i+k\eta')^2}{k^6}\biggr].
\end{equation}
Comparing the time integral appearing here with the one in~\eqref{1looptint}, one sees that they are the same after the replacement $\ve\rightarrow\delta$. Thus, there are two regions, determined by the same criteria discussed above. 

The explicit evaluation of the late-time and early time regions can be found in App.~\ref{app:PowerSpectrum}.
After adding the results for the individual regions, one recovers the sum of~\eqref{eq::oneloop2ptreg} and~\eqref{eq::dmintermediate}.
Taking the counterterm~\eqref{eq::dm} into account, one then recovers the full renormalised result~\eqref{eq::pwrren} after letting $\Lambda\rightarrow0$.
This computation confirms that the method of regions, as applied to the time integral in Sec.~\ref{treetrispec}, also extends to the loop level, as anticipated.
However, in this simple example, the momentum integral was scaleless and, moreover, factorised from the time-integral.
To determine the UV pole, it is enough to only consider the UV region of the integral, and a proper simultaneous decomposition into regions for both the time and momentum integral is not necessary.
To verify that the method of regions also holds in more general cases, we provide an explicit one-loop calculation with non-trivial dependence on the external scales of both the time and momentum integral in the next section.

\section[Late-time limit of the one-loop bispectrum of the composite operator \texorpdfstring{$\phi^2$}{Φ\^{}2}]{Late-time limit of the one-loop bispectrum of the composite operator $\phi^2$}
\label{sec::OneLoopMixing}

\begin{figure}
    \centering
    \includegraphics[width=0.42\textwidth]{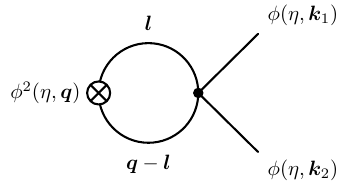}
    \caption{Topology of the one-loop, one-particle-irreducible contribution to the correlator $\cor{}{\phi^2(\eta,\vec q)\phi(\eta,\vec k_1)\phi(\eta,\vec k_2)}$. The crossed vertex denotes the insertion of $\phi^2(\eta,\vec x)$ and the dot denotes the $\kappa\phi^4$ vertex.}
    \label{fig:6:InsertionDiagram}
\end{figure}

In this section, we compute the late-time limit of the composite-operator correlator
\begin{equation}
    \cor{}{\phi^2(\eta,\vec q)\phi(\eta,\vec k_1)\phi(\eta,\vec k_2)}'
\end{equation}
at one-loop order using the method of regions. We focus on the one-particle-irreducible (1PI) contribution, which corresponds to the diagram depicted in Fig.~\ref{fig:6:InsertionDiagram}.
The motivation to study this quantity is twofold. First, it is an example of a one-loop diagram where, unlike the one-loop power spectrum considered above, the time and momentum integrals do not factorise, and the momentum integral is not scaleless.
This results in a more complicated region structure as the decomposition now involves the interplay of both time and momentum integrations.
Furthermore, the integral is sufficiently complex that an exact evaluation as a function of $k_i\eta$ before expanding in the late-time limit appears unnecessarily difficult. 
The second reason is that this diagram topology is considered in~\cite{Cohen:2021fzf} in the framework of SdSET,
where it contributes to one of the coefficients of the Kramers-Moyal equation at next-to-leading order in the power-counting.

The momentum-space correlation function is defined by Fourier transformation as
\begin{align}
&\cor{}{\phi^2(\eta,\vec q)\phi(\eta,\vec k_1)\phi(\eta,\vec k_2)}\nonumber\\
&\equiv\int\der^{d-1}x\;e^{-i\vec q\cdot\vec x}\int\biggl[\prod_{j=1}^2\frac{\der^{d-1}l_j}{(2\pi)^{d-1}}e^{i\vec l_j\cdot\vec x}\biggr]\cor{}{\phi(\eta,\vec l_1)\phi(\eta,\vec l_2)\phi(\eta,\vec k_2)\phi(\eta,\vec k_2)}\nonumber\\
&=(2\pi)^{d-1}\delta^{(d-1)}(\vec q+\vec k_1+\vec k_2)\int\frac{\der^{d-1}l}{(2\pi)^{d-1}}\cor{}{\phi(\eta,\vec l)\phi(\eta,-\vec l-\vec k_1-\vec k_2)\phi(\eta,\vec k_1)\phi(\eta,\vec k_2)}'\,,
\end{align}
where the $\delta$-function is pulled out of the correlator via the definition~\eqref{corrconvention} in the last step.
At $\Lo(\kappa)$, the momentum-space correlation function in the integrand is just the tree-level trispectrum considered in~\eqref{triint}, so one finds
\begin{align}
&\cor{}{\phi^2(\eta,\vec q)\phi(\eta,\vec k_1)\phi(\eta,\vec k_2)}'\Bigr|_{\Lo(\kappa)}\nonumber\\
&=\frac{\mu^{4-d}\kappa H^d(-H\eta)^{2d-8}}{8(k_1k_2)^3}\Im\biggl[\int_{-\infty}^{\eta}\frac{\der\eta'}{(-\eta')^{8-d}}\int\frac{\der^{d-1}l}{(2\pi)^{d-1}}\frac{e^{ik_l(\eta'-\eta)}}{l^3|\vec l-\vec q|^3}(-i+l\eta)(i+l\eta')\nonumber\\
&\hspace{0.4cm}\times(-i+|\vec l-\vec q|\eta)(i+|\vec l-\vec q|\eta')\prod_{j=1}^2(-i+k_j\eta)(i+k_j\eta')\biggr]\,,
\label{eq::22start}
\end{align}
with $k_l\equiv k_1+k_2+l+|\vec l-\vec q|$.
As we do not consider higher-order corrections to this correlator, the subscript $\Lo(\kappa)$ is omitted in the following.
This integral is quite complicated and an explicit analytical result is not known to the best of our knowledge.
For the remainder of this section, we compute the late-time limit of this expression using the method of regions.
In the individual regions, the integrand simplifies drastically and a direct evaluation is possible.

\subsection{Poles of the full result}

Before performing the region analysis, it is instructive to extract the pole terms of the expression~\eqref{eq::22start}.
These poles are related to divergences in the full integral and must be reproduced by the method of regions.
All other poles appearing in the individual regions are caused by the region decomposition and must cancel in the end when all terms are summed.
However, these additional poles give rise to the logarithms of external quantities that are generated by the expansion in the regulators, and eventually track large logarithms of the full result, which are not related to the poles of the full integral. 

After close inspection of the integrand in~\eqref{eq::22start}, one sees that the time integral is finite, but the momentum integral diverges both in the UV for $l\rightarrow\infty$ and the IR if $\vec{l}\to \vec{0}$ and $\vec{l}\to\vec{q}$.
To extract the poles, express the integrand as
\begin{equation}
\frac{g(l,|\vec l-\vec q|,k_1,k_2)}{l^3|\vec l-\vec q|^3}\,,
\end{equation}
where the numerator $g(l,|\vec l-\vec q|,k_1,k_2)$ is defined by
\begin{align}
    g(l,|\vec l-\vec q|,k_1,k_2)\equiv& e^{ik_l(\eta'-\eta)}(-i+l\eta)(i+l\eta')(-i+|\vec l-\vec q|\eta)(i+|\vec l-\vec q|\eta')\nonumber\\
    &\times\prod_{j=1}^2(-i+k_j\eta)(i+k_j\eta')\,,
\end{align}
then add and subtract the expansion of the integrand in the singular limits as
\begin{align}
\frac{g(l,|\vec l-\vec q|,k_1,k_2)}{l^3|\vec l-\vec q|^3}&=\underbrace{\biggl[\frac{g(l,|\vec l-\vec q|,k_1,k_2)}{l^3|\vec l-\vec q|^3}-\frac{g(l,l,k_1,k_2)}{l^6}\biggr|_{l\gg k_{1,2}}-2\,\frac{g(l,q,k_1,k_2)}{l^3q^3}\biggr|_{l\ll k_{1,2}}\biggr]}_{\textrm{UV- and IR-finite}}\nonumber\\
&\hspace{0.4cm}+\frac{g(l,l,k_1,k_2)}{l^6}\biggr|_{l\gg k_{1,2}}+2\,\frac{g(l,q,k_1,k_2)}{l^3q^3}\biggr|_{l\ll k_{1,2}}\,.\label{eq:6:FullPolesSubtracted}
\end{align}
As the first three terms are finite in both the UV and IR by construction, the poles are generated only by the last two summands.
Since these terms are expanded in $l\gg k_{1,2}$ and $l\ll k_{1,2}$, respectively, they are simpler to integrate than the full expression.
Moreover, the first of the two terms contains the UV pole, while the IR poles originate from the second one.
The details of this computation are presented in App.~\ref{app::poles}, and the resulting poles, with their origin indicated by the subscript, are given by
\begin{align}
&\cor{}{\phi^2(\eta,\vec q)\phi(\eta,\vec k_1)\phi(\eta,\vec k_2)}'\Bigr|_{\textrm{poles}}\nonumber\\
&=\frac{\kappa H^4}{48\pi^2(k_1k_2)^3}\biggl\{-\frac{3}{4\ve_{\textrm{UV}}}+\frac{1}{\ve_{\textrm{IR}}}\biggl[\biggl(1+\frac{k^3_1+k^3_2}{q^3}\biggr)\biggl(1-\ln\Bigl(-e^{\gamma_E}(k_1+k_2+q)\eta\Bigr)\biggr)\nonumber\\
&\hspace{3.05cm}-f(k_1,k_2,q)\biggr]\biggr\}\,,
\label{eq::22poles}
\end{align}
where 
\begin{equation}
    f(k_1,k_2,q)\equiv\frac{4k_1k_2}{q^2}-\frac{(k_1+k_2+q)[k_1(k_2+q)+k_2q]}{q^3}\,.
    \label{eq::IRf}
\end{equation}

\subsection{Decomposition of the full expression into regions}

Next, we determine the regions that contribute to the late-time limit of the correlator~\eqref{eq::22start}.
As in the previous cases, the widely separated external scales are provided by the external momenta $\vec k_{1,2}$ and the correlation time $\eta$, which satisfy
$-k_{1,2}\eta\ll1$.
Since both time and momentum are integrated over, it is natural to compare the conformal-time integration variable $\eta'$ to the correlation time $\eta$, and the modulus of the loop momentum $l$ to the moduli of the external momenta $k_{1,2}$ to identify the relevant regions.

\begin{figure}[t]
    \centering
    \includegraphics[width=0.65\textwidth]{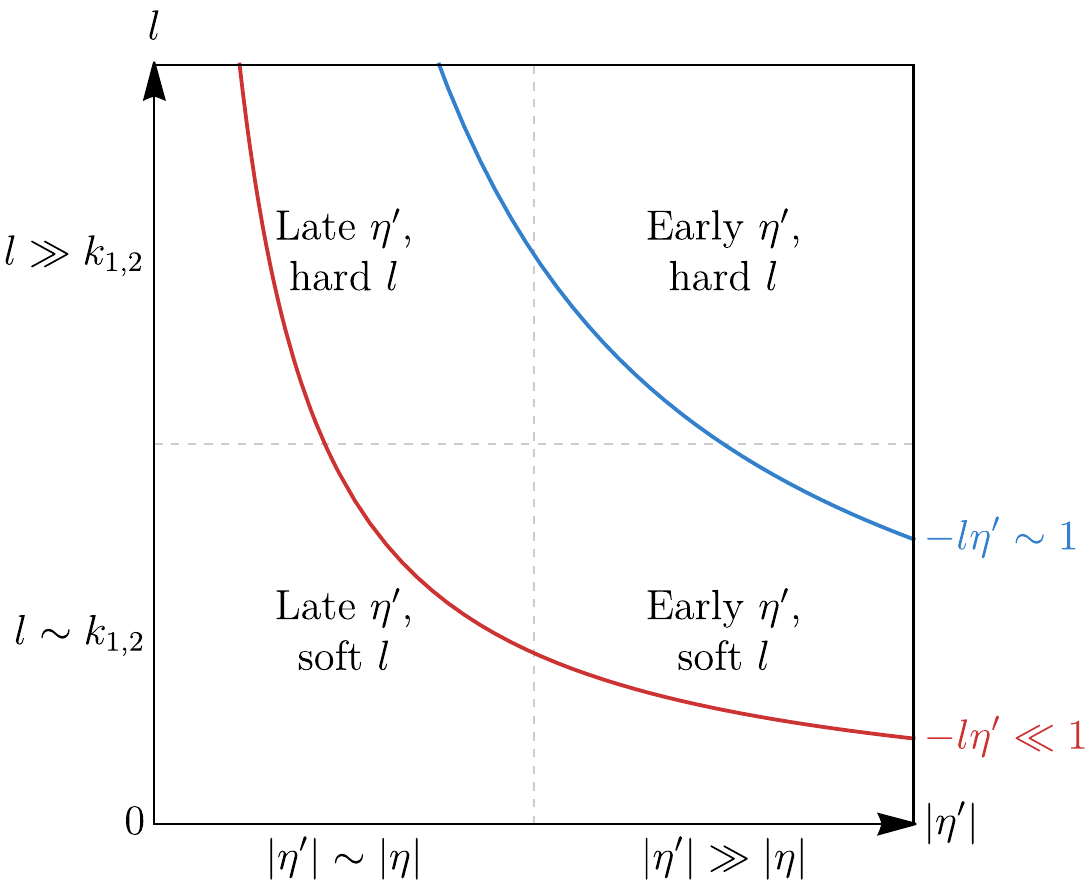}
    \caption{Graphical representation of the four regions of the double integration in~\eqref{eq::22start}. The two hyperbolae represent examples of modes which contribute to the correlation function that are superhorizon (red) or at horizon crossing (blue) at time $\eta^\prime$.}
    \label{regionsfig}
\end{figure}

The time integration can be split into two regions, determined by
\begin{equation}
    \eta'\ll\eta\quad\textrm{and}\quad \eta'\sim\eta\,,
\end{equation}
as in Sec.~\ref{triregions}, which are again called ``early-time" and ``late-time", respectively. 
Likewise, the momentum integration contains two regions, characterised by
\begin{equation}
    l\gg k_{1,2}\quad\textrm{and}\quad l\sim k_{1,2}\,,
\end{equation}
denoted as the ``hard" and ``soft" regions, respectively.
In total, one expects four different regions, as summarised in Fig.~\ref{regionsfig}. 
We organise these into two late-time and two early-time regions, distinguished by the hard and soft loop momentum.

%%%%%%%%%%%%%%%%%%%%%%%%%%%%%%%%%%%%%%%%%%%%%%%%%%%%%
%%%%%%%%%%%%%%%%%%%%%%%%%%%%%%%%%%%%%%%%%%%%%%%%%%%%%

These regions can be characterised as follows:
\begin{enumerate}[label=\arabic*)]
\item Late-time regions:
here, the conformal time $\eta^\prime$ in the integral scales as $\eta'\sim\eta$, which implies $-k_{1,2}\eta^{(\prime)}\ll1$.
The momentum integration is then split into the hard and soft regions, where the loop momentum scales as $-l\eta^{(\prime)}\sim1$ and $-l\eta^{(\prime)}\ll1$, respectively.
One performs the appropriate Taylor expansions of the integrand in the small quantities in each region while integrating over the full ranges of $\eta'$ and $\vec l$. Schematically, this gives
\begin{align}
&\int_{-\infty}^{\eta}\der\eta'\int\der^3l\biggr|_{\textrm{late time}}\nonumber\\
&=\underbrace{\int_{-\infty}^{\eta}\der\eta'\biggr|_{-k_{1,2}\eta^{(\prime)}\ll1}\int\der^3l\biggr|_{-l\eta^{(\prime)}\sim1}}_{\textrm{late $\eta'$, hard $l$}}+\underbrace{\int_{-\infty}^{\eta}\der\eta'\biggr|_{-k_{1,2}\eta^{(\prime)}\ll1}\int\der^3l\biggr|_{-l\eta^{(\prime)}\ll1}}_{\textrm{late $\eta'$, soft $l$}}\,,
\end{align}
where the integrands are understood to be expanded in the above equation.

\item Early-time regions: here, the scaling is given by $\eta'\ll\eta$, which implies $-k_{1,2}\eta'\sim1$. 
As before, the momentum integration is split into the hard and soft region, which are characterised by the scalings $-l\eta'\gg1$ and $-l\eta'\sim1$, respectively. 
In the early-time region, the integration range of $\eta'$ must be extended to $(-\infty,0)$, as explained in Sec.~\ref{4ptregions}, while $\vec l$ is integrated over its full range of values.
After expanding the integrands in the appropriate small parameters, one finds schematically
\begin{align}
&\int_{-\infty}^{\eta}\der\eta'\int\der^3l\nonumber\biggr|_{\textrm{early time}}\\
&=\underbrace{\int_{-\infty}^0\der\eta'\biggr|_{-k_{1,2}\eta'\sim1}\int\der^3l\biggr|_{-l\eta^{(\prime)}\gg1}}_{\textrm{early $\eta'$, hard $l$}}+\underbrace{\int_{-\infty}^0\der\eta'\biggr|_{-k_{1,2}\eta'\sim1}\int\der^3l\biggr|_{\substack{-l\eta'\sim1\\ -l\eta\ll1}}}_{\textrm{early $\eta'$, soft $l$}}\,.
\end{align}
\end{enumerate}

Physically, the two contributions in both the late-time and early-time regions can be associated to two different modes, distinguished by their physical momenta $-l\eta\ll1$ and $-l\eta~\sim 1$.
The first one corresponds to modes that are already superhorizon at correlation time, characterised by their small comoving momenta $l\sim k_{1,2}$.
These modes will provide the time-dependence of the secular logs and are the dynamic degrees of freedom that survive in the late-time limit $\eta\to0$.
The correlators of these modes depend on all external scales.

In the early-time region, these modes appear blue-shifted with $-l\eta^\prime\sim 1$, i.e.~the region captures their physics at the time of horizon crossing.
These modes ``freeze out" as they cross the horizon.
The effect of their self-interactions inside the horizon manifests itself in the form of a time-independent contribution to the late-time correlation function, which takes the form of a function of the external comoving momenta $\vec k_{1,2}$. In the context of SdSET, this contribution is captured by including non-Gaussian initial conditions in the definition of the EFT, which get imprinted on the initial-time slice of the theory as the field modes get redshifted and cross the horizon. 

The second type of contribution, the hard-momentum region, can be associated to modes with $-l\eta\sim 1$, i.e.~modes that are still subhorizon at correlation time.
Their effect is similar to high-energy fluctuations in usual flat-space QFT. Since they originate from modes with large momenta running in the loop, their contribution to the correlation function must be spatially local, and thus independent of $\vec k_{1,2}$. The same does not need to hold for the $\eta$-dependence, since the modes in question probe both early and late times.
However, as these modes have not yet crossed the horizon, the distinction into a late-time and early-time region is meaningless, as there is no additional physical scale to compare both regions to.
Consequently, the early-time, hard-momentum region is generically scaleless and vanishes.
In the method of regions, one can see this directly by observing that the corresponding integrands are expanded in all external scales and no scale is left inside the integral.

At this point, we must remark on a complication that arises when computing the late-time, hard-momentum and early-time, soft-momentum regions.
Both regions are characterised by physical momenta of order $-l\eta'\sim1$,
but for different reasons: for the hard region, $\eta'$ is a small number while $l$ is large, whereas in the other instance, the situation is reversed.
This can also be seen in Fig.~\ref{regionsfig}, as the hyperbola corresponding to $-l\eta'\sim1$ extends into both aforementioned regions. However, by construction, our regularisation scheme is only sensitive to physical momenta, such that the regularised integrands depend on $l$ and $\eta'$ only through the combination $-l\eta'$, which has the same scaling in both regions.
This means that the dimensional regulator is not sensitive to the difference between the two regions, resulting in integrals that are no longer regularised in the limits where the two regions overlap.
Technically, this means that dimensional regularisation alone is insufficient to properly regulate all integrals appearing in the region decomposition.\footnote{We have checked that the same situation occurs also when using an  analytic regulator of the form $(-l\eta')^{-2\ve}$ to regularise \eqref{eq::22start}. As such, the necessity of an additional regulator here has a physical motivation, in contrast to the accidental cancellation that was observed in Sec.~\ref{sec::mcounter}.}

A similar situation occurs in soft-collinear effective theory (SCET), in particular in the effective theory called SCET$_{\textrm{II}}$~\cite{SCETII,Beneke:2003pa}, which describes, among others, so-called collinear and soft modes characterised by the same virtuality but different rapidities. In diagrams involving these modes, a regularisation scheme which is sensitive only to the virtuality of the modes, such as dimensional regularisation, is not sufficient to regularise all divergent integrals and a dedicated ``rapidity regulator" must be introduced~\cite{Beneke:2003pa}. 
This regulator necessarily breaks boost invariance in order to differentiate between collinear and soft modes, but boost invariance is recovered once all contributions to a process are added back together. 
For a systematic treatment of rapidity divergences in SCET$_{\textrm{II}}$, see~e.g.~\cite{rapdiv}.

To deal with this issue in the present computation, we introduce an additional, ad-hoc analytic regulator\footnote{One could equivalently choose to regulate the momentum integrals via $(\nu/l)^{-2\delta}$, but regulating the time integral is more convenient in the following.}
\begin{equation}
    (-\nu\eta')^{-2\delta}
\end{equation}
with its own scale $\nu$ (not to be confused with the mass parameter appearing in the free mode functions, which we fixed to $3/2$), into the various regions, which then regularises the appearing integrals even in the overlap region of early-time, soft-momentum and late-time, hard-momentum.
We call this regulator the ``comoving" regulator to differentiate it from the dimensional regulator. 
After performing the integrals, the results must be expanded in the double limit $\ve,\delta\rightarrow0$, and the order in which these limits are taken must be specified.
In this case, the order is clear:
one should first take $\delta\to0$ then $\ve\to 0$, as the starting expression~\eqref{eq::22start} is fully regularised by dimensional regularisation alone, and there is no need to introduce the comoving regulator.
As such, this ordering prescription is equivalent to introducing the $\delta$-regulator in the full expression~\eqref{eq::22start} for the purpose of decomposing it into regions in a well-defined way, and simply setting it to 1 when doing the direct computation.
As in the example of the rapidity regulator in SCET$_{\textrm{II}}$, this regulator necessarily breaks the invariance of the single regions under the de Sitter isometries. However, once all regions are summed, the dependence on this regulator drops out and the invariance is again restored.

We now proceed to evaluate the four regions of the correlator~\eqref{eq::22start} explicitly, identifying the already existing and new UV- and IR-poles and showing the complete cancellation of all additional poles.
This way, we determine the leading-order asymptotic expansion of the correlator in the late-time limit by performing straightforward integrals.

\subsection{Late-time regions} 

We begin by computing the two late-time regions.
In each region, the pole structure is analysed and the origin of the poles from either UV or IR divergences is indicated.
This allows for the separation of the pole terms
into poles that are already present in the full result, and the ones that are introduced by the region decomposition.
These additional poles are of key interest, as they encode information about the factorisation of the logarithms of widely separated external scales. We further obtain the finite parts, which can be associated with correlators and matching coefficients of SdSET.

\subsubsection{Soft momentum integral}

In this region, the integrand is expanded in all of its arguments, assuming all of them to be comparably small, as in the case of the tree-level trispectrum \eqref{softtri}. There, it was observed that the only terms surviving in the late-time limit are the ones proportional to the sum over cubes of momenta.
This holds true here as well. The starting point for the computation is thus
\begin{align}
&\cor{}{\phi^2(\eta,\vec q)\phi(\eta,\vec k_1)\phi(\eta,\vec k_2)}'\Bigr|_{\textrm{late }\eta',\,\textrm{soft }l}=-\frac{\tilde\mu^{4-d}\kappa H^d(-H\eta)^{2d-8}}{24(k_1k_2)^3}\nonumber\\
&\hspace*{1cm}\times\,\nu^{-2\delta}\int_{-\infty}^{\eta}\frac{\der\eta'}{(-\eta')^{8-d+2\delta}}\Bigl[\eta^3+(-\eta')^3\Bigr]\,\int\frac{\der^{d-1}l}{(2\pi)^{d-1}}\frac{l^3+|\vec l-\vec q|^3+k_1^3+k_2^3}{l^3|\vec l-\vec q|^3}\,.
\end{align}
The dimensional regulator suffices to regularise both the time and momentum integrals, and one can set $\delta\rightarrow0$ before integration. 
The time integral can be solved directly, which results in
\begin{align}\label{eq::latetsoftlinter}
&\cor{}{\phi^2(\eta,\vec q)\phi(\eta,\vec k_1)\phi(\eta,\vec k_2)}'\Bigr|_{\textrm{late }\eta',\,\textrm{soft }l}\\
&=-\frac{\tilde\mu^{4-d}\kappa H^d(-H\eta)^{2d-8}}{8(k_1k_2)^3}\frac{(-\eta)^{-2\ve}}{2\ve(3+2\ve)}\biggl[2\int\frac{\der^{d-1}l}{(2\pi)^{d-1}}\frac{1}{l^3}+[k_1^3+k_2^3]\int\frac{\der^{d-1}l}{(2\pi)^{d-1}}\frac{1}{l^3|\vec l-\vec q|^3}\biggr]\,.\nonumber
\end{align}
Here, the first integral is scaleless and vanishes.
The second integral is UV-finite but IR-divergent, and can be solved using the standard massless-propagator integral
\begin{equation}
\tilde\mu^{2\ve}\int\frac{\der^{d-1}l}{(2\pi)^{d-1}}\frac{1}{(l^2)^{\frac{a}{2}}[(\vec l-\vec q)^2]^{\frac{b}{2}}}=\frac{(e^{\gamma_E}\mu)^{2\ve}}{8\pi^{\frac{3}{2}}}\frac{\Gamma(\frac{a+b-3}{2}+\ve)\Gamma(\frac{3-a}{2}-\ve)\Gamma(\frac{3-b}{2}-\ve)}{\Gamma(\frac{a}{2})\Gamma(\frac{b}{2})\Gamma(3-\frac{a+b}{2}-2\ve)}q^{3-a-b-2\ve}\,.
\end{equation}
The resulting expression for the correlator in the late-time, soft-momentum region reads
\begin{align}
&\cor{}{\phi^2(\eta,\vec q)\phi(\eta,\vec k_1)\phi(\eta,\vec k_2)}'\Bigr|_{\textrm{late }\eta',\,\textrm{soft }l}\nonumber\\
&=\frac{\kappa H^4(-H\eta)^{-4\ve}}{96\pi^2(k_1k_2)^3}\biggl(\frac{e^{\gamma_E}\mu}{H}\biggr)^{\!2\ve}(-2e^{\gamma_E}q\eta)^{-2\ve}\frac{k_1^3+k_2^3}{q^3}\biggl[\frac{1}{\ve^2}+\frac{4}{3\ve}+\frac{\pi^2}{12}-\frac{8}{9}\biggr]\,.
\end{align}
As anticipated, the result depends both on $\vec k_{1,2}$ and $\eta$. Furthermore, it contains both double- and single-poles in $\ve$, and is thus more divergent than the full result, which only contains single-poles in $\ve$, as seen in~\eqref{eq::22poles}. Since this region captures the infrared properties of the original integrand faithfully, the IR-poles of this region correspond to (part of) the IR-poles of the full expression. The additional poles, and in particular the double-pole, should originate from the UV. Indeed, the pole of the time integral that appears as an overall factor in~\eqref{eq::latetsoftlinter} stems from the limit $\eta'\rightarrow-\infty$, and is a UV-divergence, while the momentum integrals generate both UV- and IR-poles.
Distinguishing the poles by a subscript, we find the structure
\begin{equation}
    -\frac{\kappa H^4(-H\eta)^{-4\ve}}{96\pi^2(k_1k_2)^3}\biggl(\frac{e^{\gamma_E}\mu}{H}\biggr)^{\!2\ve}\biggl[\frac{1}{\ve_{\textrm{UV}}}-\frac{2}{3}-2\ln(-2e^{\gamma_E}q\eta)\biggr]\biggl[\frac{1}{\ve_{\textrm{UV}}}-\frac{1}{\ve_{\textrm{IR}}}-\frac{k_1^3+k_2^3}{q^3}\biggl(\frac{1}{\ve_{\textrm{IR}}}+2\biggr)\biggr]\,.\label{eq::latetsoftlpoles}
\end{equation}
When expanding this coefficient, the poles generate single- and double-logarithms of the form 
\begin{equation}
    \ln\biggl(-\frac{\mu}{2q\eta H}\biggr)=\ln\biggl(\frac{a(\eta)\mu}{2q}\biggr)\,,
    \label{eq::mulog1}
\end{equation}
with coefficients determined by the pole terms. Even though the $\ve_{\textrm{UV}}$-poles drop out when summing all regions, the above $\eta$-dependent logarithms that they generate remain and reproduce the secular logarithms of the full result. 
As was the case for the tree-level trispectrum, the $\mu$-dependence in this region tracks the $\eta$-dependence. 
The natural scale choice for $\mu$ in the late-time, soft-momentum region is $\mu\sim q/a(\eta)\ll H$, ensuring that the logarithms of the form \eqref{eq::mulog1} are small.

In the context of the SdSET, this region corresponds to the correlation function of effective fields $\cor{}{\vp^2_+(\eta,\vec q)\vp_+(\eta,\vec k_1)\vp_+(\eta,\vec k_2)}$ evaluated at one-loop order using the EFT four-point vertex. This reflects the reasoning applied above to identify the different regions, since the dynamical degrees of freedom described in the EFT with momentum $\vec k$ at an arbitrary time $\eta'$ must satisfy $-k\eta'\ll1$, which are precisely the modes contributing to the above result.

\subsubsection{Hard momentum region}

In this region, $-k_{1,2}\eta\sim-k_{1,2}\eta'\ll1$ but $-l\eta\sim-l\eta'\sim1$, so the correlator reads
\begin{align}\label{eq::22hardlsoftetastart}
&\cor{}{\phi^2(\eta,\vec q)\phi(\eta,\vec k_1)\phi(\eta,\vec k_2)}'\Bigr|_{\textrm{ late $\eta'$, hard $l$}}\\
&=\frac{\tilde\mu^{4-d}\kappa H^d(-H\eta)^{2d-8}}{8(k_1k_2)^3}\nu^{-2\delta}\Im\biggl[\int_{-\infty}^{\eta}\frac{\der \eta'}{(-\eta')^{8-d+2\delta}}\int\frac{\der^{d-1}l}{(2\pi)^{d-1}}\frac{e^{2il(\eta'-\eta)}(-i+l\eta)^2(i+l\eta')^2}{l^6}\biggr]\,.\nonumber
\end{align}
To evaluate this expression, one computes the master integral
\begin{align}
I(a,b)&\equiv\int_{-\infty}^{\eta}\frac{\der\eta'}{(-\eta')^a}\int\frac{\der^{d-1}l}{(2\pi)^{d-1}}\frac{e^{2il(\eta'-\eta)}}{l^b}\\
&=\frac{i(4\pi)^{\ve}(2i)^{b+2\ve}\Gamma(3-b-2\ve)\Gamma(b-2+2\ve)\Gamma(2+a-b-2\ve)}{32\pi^{\frac{3}{2}}\Gamma(\frac{3}{2}-\ve)\Gamma(a)}(-\eta)^{b-a-2+2\ve}\,,\nonumber
\end{align}
which satisfies the integration-by-parts relations
\begin{align}
I(a,b)&=\frac{2i\eta}{2+a-b-2\ve}I(a,b-1)\,,\\
I(a,b)&=\frac{1+a-b-2\ve}{\eta(1-a)}I(a-1,b)\,.
\end{align}
These relations can be used to reduce the correlator~\eqref{eq::22hardlsoftetastart} to a single integral, multiplied by a rational function of $\delta$, $\ve$ and $\eta$, in the form
\begin{align}
&\cor{}{\phi^2(\eta,\vec q)\phi(\eta,\vec k_1)\phi(\eta,\vec k_2)}'\Bigr|_{\textrm{ late $\eta'$, hard $l$}}\nonumber\\
&=\frac{\tilde\mu^{4-d}\kappa H^d(-H\eta)^{2d-8}}{8(k_1k_2)^3}\nu^{-2\delta}\frac{(4+2\delta)[2(\delta+\ve)-1]\eta^2}{2\delta[3+2(\delta+\ve)]}I\Bigl(2+2(\delta+\ve),2\Bigr)\nonumber\\
&=\frac{\kappa H^4(-H\eta)^{-4\ve}}{96\pi^2(k_1k_2)^3}\biggl(\frac{e^{\gamma_E}\mu}{H}\biggr)^{\!2\ve}(-\nu\eta)^{-2\delta}\biggl[\frac{1}{\delta}\biggl(\frac{1}{\ve}-\frac{8}{3}\biggr)-\frac{13}{6\ve}+\frac{56}{9}-\frac{2\pi^2}{3}\biggr]\,.
\end{align}
The result of the loop integral is independent of $\vec k_{1,2}$ and is therefore spatially local, as expected, but it has a non-trivial dependence on $\eta$. The contribution from this region can thus be written as the tree-level correlator $\cor{}{\phi^2(\eta,\vec q)\phi(\eta,\vec k_1)\phi(\eta,\vec k_2)}'$ multiplied by a time-dependent coefficient containing poles. 

As in the previous subsection, we identify the origin of the poles. When extracting the UV-pole of the full expression as in~\eqref{eq:6:FullPolesSubtracted}, one expands the full integrand in a very similar fashion to~\eqref{eq::22hardlsoftetastart} 
(see also~\eqref{eq::22UVpolestart}).
The main difference is that in the full result, one keeps the terms proportional to $k_{1,2}$. 
However, as the UV-pole is independent of $k_{1,2}$, this is irrelevant.
So the UV-pole of the full expression must be the same as the UV-pole of this region.
Consequently, the remaining poles must originate from IR divergences. Separating these poles, one then finds
\begin{align}\label{eq::latethardlpoles}
    &\frac{\kappa H^4(-H\eta)^{-4\ve}}{96\pi^2(k_1k_2)^3}\biggl(\frac{e^{\gamma_E}\mu}{H}\biggr)^{\!2\ve}\biggl\{\biggl[\frac{1}{\delta}-2\ln(-\nu\eta)\biggr]\biggl[\frac{1}{\ve}-\frac{8}{3}\biggr]-\frac{13}{6\ve}\biggr\}\\
    &=\frac{\kappa H^4(-H\eta)^{-4\ve}}{96\pi^2(k_1k_2)^3}\biggl(\frac{e^{\gamma_E}\mu}{H}\biggr)^{\!2\ve}\biggl\{\frac{1}{\delta_{\textrm{IR}}\ve_{\textrm{IR}}}-\frac{3}{2\ve_{\textrm{UV}}}-\frac{8}{3\delta_{\textrm{IR}}}-\frac{2}{\ve_{\textrm{IR}}}\biggl[\frac{1}{3}+\ln(-\nu\eta)\biggr]+\Lo(\delta^0,\ve^0)\biggr\}\,.\nonumber
\end{align}
Here, the $\delta$-expansion generates single logarithms of the form 
\begin{equation}
    \ln(-\nu\eta)=\ln\biggl(\frac{\nu}{a(\eta)H}\biggr)
    \label{eq::nulog1}
\end{equation} 
while the $\ve$-expansion generates logarithms of the form
\begin{equation}
    \ln\biggl(\frac{e^{\gamma_E}\mu}{H}\biggr)\,.
    \label{eq::mulog2}
\end{equation}
Since the poles in $\delta$ must cancel in the sum of all regions, so must the $\nu$-dependence, which we verify explicitly below. 
However, when considering individual regions, one can use the scale $\nu$ to control the logarithms.
In the late-time, hard-momentum region, the logarithms of the form~\eqref{eq::nulog1} can be rendered small by choosing~$\nu\sim -\eta^{-1}$. 
The same reasoning applies to the $\mu$-dependent logarithms of the form~\eqref{eq::mulog2}, which are small when choosing $\mu\sim H$. This corresponds to the natural scale in this region for $\mu$ and $\nu$.

The modes contributing to this region have momenta $l\gg k_{1,2}$ and therefore lie outside the regime of validity of SdSET. When reproducing this computation in the context of the EFT, the result of this region contains new information that must be included in the effective theory through a matching coefficient.

\subsection{Early-time regions}

\subsubsection{Soft-momentum region}
In this region, $-k_{1,2}\eta\ll-k_{1,2}\eta'\sim1$, $-l\eta\ll-l\eta'\sim1$, and the expanded correlator reads
\begin{align}
&\cor{}{\phi^2(\eta,\vec q)\phi(\eta,\vec k_1)\phi(\eta,\vec k_2)}'\Bigr|_{\textrm{ early $\eta'$, soft $l$}}\nonumber\\
&=\frac{\kappa \tilde\mu^{4-d} H^d(-H\eta)^{2d-8}}{8(k_1k_2)^3}\nu^{-2\delta}\Im\biggl[\int_{-\infty}^0\frac{\der \eta'}{(-\eta')^{8-d+2\delta}}\int\frac{\der^{d-1}l}{(2\pi)^{d-1}}\;\frac{e^{ik_l\eta'}(i+l\eta')(i+|\vec l-\vec q|\eta')}{l^3|\vec l-\vec q|^3}\nonumber\\
&\quad\times\prod_{j=1}^2(i+k_j\eta')\biggr]\,.
\label{eq::22earlytsoftlstart}
\end{align}
The appearing integrals can be reduced by defining the ``master integral"
\begin{equation}
I(a)\equiv\int_{-\infty}^0\frac{\der\eta'}{(-\eta')^{8-d+a}}e^{iK\eta'}\int\frac{\der^{d-1}l}{(2\pi)^{d-1}}\frac{e^{i(l+|\vec l-\vec q|)\eta'}(i+l\eta')(i+|\vec l-\vec q|\eta')}{l^3|\vec l-\vec q|^3}\,,
\end{equation}
with $K\equiv k_1+k_2$, which is evaluated in App.~\ref{app::int}.
The integral $I(a)$ satisfies the differential equation
\begin{equation}
\frac{\p}{\p K}I(a)=-iI(a-1)\,,
\end{equation}
and therefore the correlator~\eqref{eq::22earlytsoftlstart} can be expressed in terms of a differential operator acting on the single integral $I(2\delta)$, as
\begin{align}
&\hspace{-0.4cm}\cor{}{\phi^2(\eta,\vec q)\phi(\eta,\vec k_1)\phi(\eta,\vec k_2)}'\Bigr|_{\textrm{ early $\eta'$, soft $l$}}\nonumber\\
&=\frac{\kappa \tilde\mu^{4-d} H^d(-H\eta)^{2d-8}}{8(k_1k_2)^3}\nu^{-2\delta}\Im\biggl[\biggl(-1+K\frac{\p}{\p K}-k_1k_2\frac{\p^2}{\p K^2}\biggr)I(2\delta)\biggr]\biggr|_{K=k_1+k_2}\nonumber\\
&=\frac{H^4(-H\eta)^{-4\ve}}{96\pi^2(k_1k_2)^3}\biggl(\frac{e^{\gamma_E}\mu}{H}\biggr)^{\!2\ve}\biggl(\frac{2e^{\gamma_E}q}{\nu}\biggr)^{\!2\delta}\biggl\{-\frac{1}{\ve^2}\frac{k_1^3+k_2^3}{q^3}-\frac{1}{\delta}\biggl[\frac{1}{\ve}-\frac{8}{3}\biggr]\nonumber\\
&+\frac{2}{\ve}\biggl[\biggl(1+\frac{k_1^3+k_2^3}{q^3}\biggr)\biggl(\frac{1}{3}-\ln\biggl(\frac{k_1+k_2+q}{2q}\biggr)\biggr)+1-f(k_1,k_2,q)\biggr]\nonumber\\
&+\frac{\pi^2}{12}\biggl[14-\frac{k_1^3+k_2^3}{q^3}\biggr]-\frac{4}{9}\biggl[26+\frac{12(k_1+k_2)}{q}+\frac{3(4k_1^2-k_1k_2+4k_2^2)}{q^2}+\frac{(k_1+k_2)^3}{q^3}\biggr]\nonumber\\
&+\frac{4}{3}\biggl[4+\frac{3(k_1+k_2)}{q}+\frac{(k_1+k_2)^3}{q^3}\biggr]\ln\biggl(\frac{k_1+k_2+q}{2q}\biggr)\nonumber\\
&-\biggl[\biggl(1+\frac{k_1^3+k_2^3}{q^3}\biggr)\ln^2\biggl(\frac{k_1+k_2+q}{2q}\biggr)\nonumber\\
&\hspace{0.7cm}+\biggl(1-\frac{k_1^3+k_2^3}{q^3}\biggr)\ln\biggl(\frac{k_1+k_2}{q}\biggr)\ln\biggl(\frac{4(k_1+k_2)q}{(k_1+k_2-q)^2}\biggr)\biggr]\nonumber\\
&+2\biggl(\frac{k_1^3+k_2^3}{q^3}-1\biggr)\biggl[\Li_2\biggl(\frac{q}{k_1+k_2}\biggr)-\Li_2\biggl(-\frac{q}{k_1+k_2}\biggr)+\Li_2\biggl(\frac{1}{2}-\frac{q}{2(k_1+k_2)}\biggr)\biggr]\biggr\}\,,
\end{align}
with $f(k_1,k_2,q)$ defined in \eqref{eq::IRf}.
Note that the integral $I(2\delta)$, and correspondingly also the (poly-)logarithms depend on $k_1$, $k_2$ only through the sum $K=k_1+k_2$ (see also~\eqref{eq:App:I2delta}). The entire non-trivial dependence on $k_1, k_2$ enters due to the second-derivative term explicitly proportional to the product $k_1 k_2$.

For the identification of the poles in this region, it is most convenient to evaluate the time integrals in \eqref{eq::22earlytsoftlstart} before performing the $l$-integration, and then inspect the behaviour of the resulting integrand for large and small $l$. 
The time integral yields an IR-divergence which stems from the upper integration boundary $\eta'\rightarrow0$, while the $l$-integral contributes both UV- and IR-divergences, originating from $l\rightarrow\infty$ and $l\rightarrow0$, $\vec l\rightarrow\vec q$, respectively. 
The additional analytic regulator becomes necessary to regularise the $l\rightarrow\infty$ limit, such that the $\delta$-poles can be identified as UV-poles. Separating all poles, we find
\begin{align}
    &\frac{H^4(-H\eta)^{-4\ve}}{96\pi^2(k_1k_2)^3}\biggl(\frac{e^{\gamma_E}\mu}{H}\biggr)^{\!2\ve}\Biggl\{-\frac{1}{\ve^2}\frac{k_1^3+k_2^3}{q^3}-\biggl[\frac{1}{\delta}+2\ln\biggl(\frac{2e^{\gamma_E}q}{\nu}\biggr)\biggr]\biggl[\frac{1}{\ve}-\frac{8}{3}\biggr]\nonumber\\
    &\hspace{0.5cm}+\frac{2}{\ve}\biggl[\biggl(1+\frac{k_1^3+k_2^3}{q^3}\biggr)\,\biggl(\frac{1}{3}-\ln\biggl(\frac{k_1+k_2+q}{2q}\biggr)\biggr)+1-f(k_1,k_2,q)\biggr]\Biggr\}
    \label{eq::earlytsoftlpoles}\\
    &=\frac{H^4(-H\eta)^{-4\ve}}{96\pi^2(k_1k_2)^3}\biggl(\frac{e^{\gamma_E}\mu}{H}\biggr)^{\!2\ve}\Biggl\{\frac{1}{\ve_{\textrm{IR}}}\biggl[\frac{1}{\ve_{\textrm{UV}}}-\frac{1}{\ve_{\textrm{IR}}}\biggl(1+\frac{k_1^3+k_2^3}{q^3}\biggr)\biggr]-\frac{1}{\delta_{\textrm{UV}}\ve_{\textrm{IR}}}+\frac{8}{3\delta_{\textrm{UV}}}\nonumber\\
    &\hspace{0.5cm}-\frac{2}{\ve_{\textrm{IR}}}\biggl[\frac{k_1^3+k_2^3}{q^3}+\ln\biggl(\frac{2e^{\gamma_E}q}{\nu}\biggr)\biggr]\nonumber\\
    &\hspace{0.5cm}+\frac{2}{\ve_{\textrm{IR}}}\biggl[\biggl(1+\frac{k_1^3+k_2^3}{q^3}\biggr)\biggl(\frac{4}{3}-\ln\biggl(\frac{k_1+k_2+q}{2q}\biggr)\biggr)-f(k_1,k_2,q)\biggr]+\Lo(\delta^0,\ve^0)\Biggr\}\nonumber
\end{align}
Here the $\delta$-expansion generates single logarithms of the form 
\begin{equation}
    \ln\biggl(\frac{2e^{\gamma_E}q}{\nu}\biggr)
    \label{eq::nulog2}
\end{equation}
while the $\ve$-expansion generates single- and double-logarithms of the form
\begin{equation}
    \ln\biggl(\frac{e^{\gamma_E}\mu}{H}\biggr)\,.
\end{equation}
The logarithms can be rendered small by the natural choices $\nu\sim q$ corresponding to the soft comoving momentum and $\mu\sim H$ the hard scale of the early-time region as observed also in~\eqref{hardtri}. 

In the context of SdSET, this region contributes to $\cor{}{\vp^2_+(\eta,\vec q)\vp_+(\eta,\vec k_1)\vp_+(\eta,\vec k_2)}$ at one-loop through the insertion of the tree-level, four-point, non-Gaussian initial condition. This matches our understanding of the modes contributing to this region since the effect of the subhorizon evolution of the now-superhorizon degrees of freedom at early times is described precisely through these initial conditions.

\subsubsection{Hard-momentum region}

This final region is characterised by the scalings $-k_{1,2}\eta\ll-k_{1,2}\eta'\sim1$, $-l\eta\ll-l\eta'\sim1$, and we expect this contribution to be scaleless. 
The correlator reads
\begin{align}
&\cor{}{\phi^2(\eta,\vec q)\phi(\eta,\vec k_1)\phi(\eta,\vec k_2)}'\Bigr|_{\textrm{ early $\eta'$, hard $l$}}\label{earlyetahardl}\\
&=\frac{\kappa \tilde\mu^{4-d} H^d(-H\eta)^{2d-8}}{8(k_1k_2)^3}\nu^{-2\delta}\,\Im\biggl[\int\frac{\der^{d-1}l}{(2\pi)^{d-1}}\int_{-\infty}^0\frac{\der \eta'}{(-\eta')^{8-d+2\delta}}\;\frac{e^{2il\eta'}(i+l\eta')^2}{l^6}\prod_{j=1}^2(i+k_j\eta')\biggr]\,,\nonumber
\end{align}
and contains integrals of the form
\begin{equation}
I(a,b)=\int\frac{\der^{d-1}l}{(2\pi)^{d-1}}\int_{-\infty}^0\frac{\der\eta'}{(-\eta')^a}\;\frac{e^{2il\eta'}}{l^b}\,.
\end{equation}
The time integral can be evaluated directly and reads
\begin{equation}
\int_{-\infty}^0\frac{\der\eta'}{(-\eta')^a}\;e^{2il\eta'}=(2il)^{a-1}\Gamma(1-a)\,,
\end{equation}
which leaves the momentum integral
\begin{equation}
I(a,b)=(2i)^{a-1}\Gamma(1-a)\int\frac{\der^{d-1}l}{(2\pi)^{d-1}}\frac{1}{l^{b-a+1}}\,.
\label{scaleless}
\end{equation}
This integral is scaleless and vanishes by the analytic regulator. 
Therefore, this region simply has a vanishing contribution. 
As previously discussed, this is expected, since in this region all the integration variables are taken to be much larger than the corresponding external scales, and one ends up with a scaleless expression.

However, to correctly interpret the poles induced by the region decomposition, it is important to determine whether any of these scaleless integrals contain logarithmic divergences. 
To this end, one introduces an intermediate scale $\Lambda$ and writes
\begin{equation}
I(a,b)=(2i)^{a-1}\Gamma(1-a)\frac{\Omega_{d-2}}{(2\pi)^{d-1}}\biggl[\int_0^{\Lambda}\der l\;l^{d-2+a-b-1}+\int_{\Lambda}^{\infty}\der l\;l^{d-2+a-b-1}\biggr]\,,
\label{eq::lambdaint}
\end{equation}
where the analytic regulator acts as an IR-regulator in the first integral and as a UV-regulator in the second one.
We differentiate these (and the associated scale $\nu$) using a subscript in the following.
Using the modified integral~\eqref{eq::lambdaint} in~\eqref{earlyetahardl}, one then finds
\begin{align}
&\cor{}{\phi^2(\eta,\vec q)\phi(\eta,\vec k_1)\phi(\eta,\vec k_2)}'\Bigr|_{\textrm{ early $\eta'$, hard $l$}}\nonumber\\
&=-\frac{\kappa H^4(-H\eta)^{-4\ve}}{96\pi^2(k_1k_2)^3}\biggl(\frac{e^{\gamma_E}\mu}{H}\biggr)^{\!2\ve}\biggl(\frac{\nu_{\textrm{UV}}^{-2\delta_{\textrm{UV}}}}{\delta_{\textrm{UV}}}-\frac{\nu_{\textrm{IR}}^{-2\delta_{\textrm{IR}}}}{\delta_{\textrm{IR}}}\biggr)\biggl[\frac{1}{\ve_{\textrm{IR}}}-\frac{8}{3}\biggr]\,,
\label{eq::earlythardlpoles}
\end{align}
which of course still vanishes after identifying $\delta_{\textrm{UV}}=\delta_{\textrm{IR}}$ and $\nu_{\textrm{UV}}=\nu_{\textrm{IR}}$. 
Here, the pole in $\ve$ stems from the time integral, while the ones in $\delta$ originate from the momentum integral.
From this form, one can recognise that, indeed, the integral contains logarithmic UV- and IR-divergences, which cancel. 
Note that the pole structure in~\eqref{eq::earlythardlpoles}, $\frac{1}{\delta_{\rm IR,UV}} \times (\frac{1}{\ve_{\textrm{IR}}}-\frac{8}{3})$, precisely matches the form of the $\delta$-poles in the late-time, hard-momentum region \eqref{eq::latethardlpoles} and the early-time, soft-momentum region \eqref{eq::earlytsoftlpoles}, respectively.
% \MBcomment{Note and discuss that the $\frac{1}{\delta_{\rm IV,IR}} \times (\frac{1}{\ve_{\textrm{IR}}}-\frac{8}{3})$ precisely matches the delta in  poles in \eqref{eq::latethardlpoles} and \eqref{eq::earlytsoftlpoles}.}

\subsection{Sum of the regions}
After summing all three non-vanishing regions, one obtains the final result
\begin{align}
&\hspace{-0.4cm}\cor{}{\phi^2(\eta,\vec q)\phi(\eta,\vec k_1)\phi(\eta,\vec k_2)}'\Bigr|_{\textrm{late $\eta'$, soft $l$}}+\cor{}{\phi^2(\eta,\vec q)\phi(\eta,\vec k_1)\phi(\eta,\vec k_2)}'\Bigr|_{\textrm{late $\eta'$, hard $l$}}\nonumber\\
&\hspace{-0.4cm}+\cor{}{\phi^2(\eta,\vec q)\phi(\eta,\vec k_1)\phi(\eta,\vec k_2)}'\Bigr|_{\textrm{early $\eta'$, soft $l$}}+\cor{}{\phi^2(\eta,\vec q)\phi(\eta,\vec k_1)\phi(\eta,\vec k_2)}'\Bigr|_{\textrm{early $\eta'$, hard $l$}}\nonumber\\
&=\frac{H^4(-H\eta)^{-4\ve}}{96\pi^2(k_1k_2)^3}\biggl(\frac{e^{\gamma_E}\mu}{H}\biggr)^{\!2\ve}\nonumber\\
&\times\Biggl\{\frac{2}{\ve}\biggl[\biggl(1+\frac{k^3_1+k^3_2}{q^3}\biggr)\biggl(1-\ln\Bigl(-e^{\gamma_E}(k_1+k_2+q)\eta\Bigr)\biggr)-f(k_1,k_2,q)-\frac{3}{4}\biggr]\nonumber\\
&+\frac{8}{3}\biggl[2-\frac{k_1^3+k_2^3}{q^3}\biggr]\ln(-2e^{\gamma_E}q\eta)+2\frac{k_1^3+k_2^3}{q^3}\ln^2(-2e^{\gamma_E}q\eta)\nonumber\\
&+\frac{\pi^2}{2}-\frac{4}{9}\biggl[12+\frac{12(k_1+k_2)}{q}+\frac{3(4k_1^2-k_1k_2+4k_2^2)}{q^2}+3\frac{k_1^3+k_1^2k_2+k_1k_2^2+k_2^3}{q^3}\biggr]\nonumber\\
&+\frac{4}{3}\biggl[4+\frac{3(k_1+k_2)}{q}+\frac{(k_1+k_2)^3}{q^3}\biggr]\ln\biggl(\frac{k_1+k_2+q}{2q}\biggr)\nonumber\\
&-\biggl[\biggl(1+\frac{k_1^3+k_2^3}{q^3}\biggr)\ln^2\biggl(\frac{k_1+k_2+q}{2q}\biggr)\nonumber\\
&\hspace*{0.7cm}+\biggl(1-\frac{k_1^3+k_2^3}{q^3}\biggr)\ln\biggl(\frac{k_1+k_2}{q}\biggr)\ln\biggl(\frac{4(k_1+k_2)q}{(k_1+k_2-q)^2}\biggr)\biggr]\nonumber\\
&+2\biggl(\frac{k_1^3+k_2^3}{q^3}-1\biggr)\biggl[\Li_2\biggl(\frac{q}{k_1+k_2}\biggr)-\Li_2\biggl(-\frac{q}{k_1+k_2}\biggr)+\Li_2\biggl(\frac{1}{2}-\frac{q}{2(k_1+k_2)}\biggr)\biggr]\Biggr\}\,.
\end{align}
Notably, all double poles ($\ve^2$ and $\delta\ve$), as well as the single poles in $\delta$ cancel, as they must, and the dependence on $\nu$ disappears. The remaining $\ve$-poles completely reproduce the poles of the full correlator determined in \eqref{eq::22poles}.

By collecting all pole terms obtained in \eqref{eq::latetsoftlpoles}, \eqref{eq::latethardlpoles}, and \eqref{eq::earlytsoftlpoles}, and highlighting by a subscript the poles that reproduce the ones of the full expression \eqref{eq::22poles}, we can track how the individual poles cancel between the regions:
\begin{itemize}
    \item Late-time, soft-momentum:
    \begin{align}
        &\frac{\kappa H^4(-H\eta)^{-4\ve}}{96\pi^2(k_1k_2)^3}\biggl(\frac{e^{\gamma_E}\mu}{H}\biggr)^{\!2\ve}\biggl\{\frac{1}{\ve^2}\frac{k_1^3+k_2^3}{q^3}+\frac{2}{\ve}\biggl[\frac{1}{3}+\frac{k_1^3+k_2^3}{q^3}+\ln(-2e^{\gamma_E}q\eta)\biggr]\nonumber\\
        &\quad-\frac{2}{\ve_{\textrm{IR}}}\biggl[\frac{1}{3}+\ln(-2e^{\gamma_E}q\eta)\biggr]\biggl[1+\frac{k_1^3+k_2^3}{q^3}\biggr]\biggr\}\,,\label{eq::polerecap1}
    \end{align}
    \item Late-time, hard-momentum:
    \begin{equation}
        \frac{\kappa H^4(-H\eta)^{-4\ve}}{96\pi^2(k_1k_2)^3}\biggl(\frac{e^{\gamma_E}\mu}{H}\biggr)^{\!2\ve}\biggl\{\frac{1}{\delta\ve}-\frac{3}{2\ve_{\textrm{UV}}}-\frac{8}{3\delta}-\frac{2}{\ve}\biggl[\frac{1}{3}+\ln(-\nu\eta)\biggr]\biggr\}\,,
        \label{eq::polerecap2}
    \end{equation}
    \item Early-time, soft-momentum:
    \begin{align}
        &\frac{H^4(-H\eta)^{-4\ve}}{96\pi^2(k_1k_2)^3}\biggl(\frac{e^{\gamma_E}\mu}{H}\biggr)^{\!2\ve}\biggl\{-\frac{1}{\ve^2}\frac{k_1^3+k_2^3}{q^3}-\frac{1}{\delta\ve}+\frac{8}{3\delta}-\frac{2}{\ve}\biggl[\frac{k_1^3+k_2^3}{q^3}+\ln\biggl(\frac{2e^{\gamma_E}q}{\nu}\biggr)\biggr]\nonumber\\
        &\quad+\frac{2}{\ve_{\textrm{IR}}}\biggl[\biggl(1+\frac{k_1^3+k_2^3}{q^3}\biggr)\biggl(\frac{4}{3}-\ln\biggl(\frac{k_1+k_2+q}{2q}\biggr)\biggr)-f(k_1,k_2,q)\biggr]\biggr\}\,.
        \label{eq::polerecap3}
    \end{align}
\end{itemize}
The double-$\ve$-poles cancel, as the pole from the time integral in the late-time, soft-momen\-tum region is a UV-pole, while in the early-time, soft-momentum region the time integral generates an IR-pole, and both poles have opposite sign. 
The same reasoning applies to the $\delta$-poles in \eqref{eq::polerecap2} and \eqref{eq::polerecap3}. 
Furthermore, the sum of the single-$\ve$-poles in \eqref{eq::polerecap2} and in the first line of \eqref{eq::polerecap3}, where the $\nu$-dependence drops out, is cancelled by the single-$\ve$-pole in the first line of \eqref{eq::polerecap1}.
The remaining $\ve_{\textrm{UV}}$-pole in \eqref{eq::polerecap2} and $\ve_{\textrm{IR}}$-poles in \eqref{eq::polerecap1}, \eqref{eq::polerecap3} reproduce the poles of the full expression.

In this section, we have provided an example for how the method of regions can be applied to simplify the computation of the late-time limit of non-trivial loop diagrams in de Sitter space.
The evaluation of the full, rather complicated integral~\eqref{eq::22start} could be reduced to three distinct classes of integrals, only one of which was truly non-trivial, but still significantly simpler than the starting expression. 
As an added benefit to the technical simplification, the decomposition into regions highlights which modes of the scalar field contribute to the final result. 
This provides a more transparent picture for the superhorizon physics determining the correlation function and, in the present example, allows making a clear connection between the regions and the various contributions one would encounter in the analogous computation in SdSET. 

\section{Conclusions}
\label{sec:conclusion}

The method of regions provides a way to obtain the asymptotic expansion of an integral without computing the full analytic expression.
In this article, we applied this method for the first time to cosmological in-in correlation functions of a massless, minimally-coupled scalar field in a rigid de Sitter space-time
for three explicit examples: the tree-level trispectrum, the one-loop power-spectrum and the $\phi^2$-operator insertion.
While the method works just the same as in flat space, de Sitter space comes with novel features.

As each vertex is accompanied by a time integral, the in-in correlators feature a non-trivial region structure already at tree-level,
where the correlator factorises into a time-dependent late-time contribution, corresponding to the superhorizon dynamics, and a time-independent early-time region
dominated by contributions at the time of horizon crossing.
In the context of Soft de Sitter effective theory, this points towards a non-trivial factorisation into initial conditions and superhorizon-mode correlation functions already at tree level.

For the same reason, the region structure is richer than a corresponding flat-space integral also at the one-loop level.
Here, both the time and momentum integral must be decomposed, resulting in a total of four regions,
the late-time, soft-momentum region, corresponding to modes that are superhorizon at correlation time, 
and a corresponding early-time, soft-momentum region, which forms the initial conditions imprinted at horizon crossing.
In addition, there is a late-time, hard-momentum region stemming from subhorizon modes that are still well-inside the horizon at correlation time, and a corresponding early-time, hard-momentum region.
As these modes have not yet crossed the horizon, this early-time, hard-momentum region is generically scaleless and vanishes.
Consequently, there are three distinct regions at the one-loop level. The method of regions turns the intuitive understanding of the relevant scales and contributions into a precise computational method that is in principle extensible to any loop order.

 Dimensional regularisation with an evanescent mass term provides a convenient de Sitter-invariant regulator for the four-dimensional interacting massless scalar in the full theory. In the region decomposition, however, the dimensional regulator does not distinguish the late-time hard and early-time soft regions, as both have the same physical momenta $-k\eta'\sim 1$.
This motivated the introduction of an ad-hoc analytic regulator $(-\nu\eta')^{-2\delta}$ which breaks de Sitter invariance but distinguishes both modes.
This regulator is similar in spirit to the rapidity regulator in soft-collinear effective theory, and provokes the question if similar renormalisation-group techniques can be exploited for de Sitter correlators. However, this investigation is beyond the scope of this work.

In all examples, we explicitly checked that the method of regions expansion faithfully reproduces the full result.
In the third case, the operator insertion $\langle\phi^2(\eta,\vec{q})\phi(\eta,\vec{k}_1)\phi(\eta,\vec{k}_2)\rangle$, the method of regions even allows for a straightforward evaluation of the appearing integrals including the finite terms, whereas a direct computation is rather complex.
Here, we explicitly checked that the poles of the full result are completely reproduced.

We believe that our work provides a useful starting point for the systematic investigation of cosmological correlators, particularly the secular logarithms, from an EFT perspective.
The method of regions illustrates the factorisation of these correlators, and how the different ingredients entering the effective theory can be computed in the most economic fashion.
This greatly simplifies matching computations, to which we hope to return in subsequent work.

\subsubsection*{Acknowledgement}
We thank Tim Cohen and Ivo Sachs for important discussions, and Dan Green for valuable comments on SdSET. 
This work has been supported in part by the Excellence Cluster ORIGINS funded by the Deutsche Forschungsgemeinschaft under Grant No.~EXC - 2094 - 390783311 and by the Cluster of Excellence Precision Physics, Fundamental Interactions, and Structure of Matter (PRISMA$^+$ EXC 2118/1) funded by the German Research Foundation (DFG) within the German Excellence Strategy (Project ID 390831469).

\appendix
\section{Schwinger-Keldysh diagrammatic rules}
\label{SKrules}

In the Schwinger-Keldysh formalism, each field type appearing in the path integral is doubled, with one copy of the field propagating forward in time from the initial (conformal) time $\eta_{\textrm{in}}$ to the correlation time $\eta$, and the other propagating backward in time from $\eta$ to $\eta_{\textrm{in}}$. The two copies of each field type are identified at the correlation time.  This construction effectively leads to an action in the path integral where every term, and in particular every interaction vertex, is doubled, where one copy comes with a factor of $+i$, while the other comes with a factor $-i$. We refer to these as $(+)$- and $(-)$-vertices, respectively. When drawing a diagram with $n$ vertices, one must sum over all $2^n$ versions of that diagram, where each vertex appears as a $(+)$- and $(-)$-vertex. Each vertex time in integrated from the initial-time slice, here at $\eta_{\textrm{in}}=-\infty$, to the correlation time $\eta$. 
The doubling of the vertices requires to define four types of (momentum-space) propagators
\begin{equation}\label{ining}
\begin{aligned}
G_{++}(\eta,\vec k;\eta',\vec k')&\equiv\cor{}{T\{\phi(\eta,\vec k)\phi(\eta' ,\vec k')\}}\,,\\
G_{+-}(\eta,\vec k;\eta',\vec k')&\equiv\cor{}{\phi(\eta',\vec k')\phi(\eta,\vec k)}\,,\\
G_{-+}(\eta,\vec k;\eta',\vec k')&\equiv\cor{}{\phi(\eta,\vec k)\phi(\eta',\vec k')}\,,\\
G_{--}(\eta,\vec k;\eta',\vec k')&\equiv\cor{}{\overline{T}\{\phi(\eta,\vec k)\phi(\eta',\vec k')\}}\,,
\end{aligned}
\end{equation}
where $\overline{T}$ and $T$ denote the (anti-)time-ordering operator. The (anti-)time-ordered two-point functions can be written as
\begin{equation}
\begin{aligned}
    G_{++}(\eta,\vec k;\eta',\vec k')&=\theta(\eta-\eta')\cor{}{\phi(\eta,\vec k)\phi(\eta',\vec k')}+\theta(\eta'-\eta)\cor{}{\phi(\eta',\vec k')\phi(\eta,\vec k)}\\
    G_{--}(\eta,\vec k;\eta',\vec k')&=\theta(\eta-\eta')\cor{}{\phi(\eta',\vec k')\phi(\eta,\vec k)}+\theta(\eta'-\eta)\cor{}{\phi(\eta,\vec k)\phi(\eta',\vec k')}\,,
\end{aligned}
\end{equation}
such that all four types of propagators can be expressed by means of the Wightman function $\cor{}{\phi(\eta,\vec k)\phi(\eta',\vec k')}$ and its complex conjugate.

The diagrammatic rules for this formalism can be found e.g. in the appendix of \cite{Weinberg:2005vy} and we adapt them to our notation here:
\begin{itemize}
    \item When connecting an external line to a $(+)$- or a $(-)$-vertex, use $G_{++}$ or $G_{--}$, respectively, and take into account that all vertex times $\eta_n$ in a diagram satisfy $\eta\geq \eta_n$. Because the time-ordering is fixed uniquely by this, the propagators attached to external lines always reduce to single Wightman functions.
    \item When connecting internal $(+)$- or $(-)$-vertices to each other, use the appropriate propagator from the list above, taking into account which field is associated with which vertex. For example, if the field $\phi(\eta_1,\vec k_1)$ is associated with a $(+)$-vertex and the field $\phi(\eta_2,\vec k_2)$ with a $(-)$-vertex, they are connected using $G_{+-}(\eta_1,\vec k_1;\eta_2,\vec k_2)$.
\end{itemize}
These rules ensure that the final result for an in-in correlation function is a real-valued function \cite{SKdiagrammatics}.

\section{Results for the one-loop power spectrum}\label{app:PowerSpectrum}

This appendix contains the explicit results for the late-time and early-time regions of the one-loop power spectrum discussed in Sec.~\ref{sec:OneLoopRegions}.

In the late-time region $\eta^\prime\sim\eta$, corresponding to $-k\eta^\prime\to0$, the power spectrum reads
\begin{align}
&\cor{}{\phi(t,\vec k)\phi(t,-\vec k)}'\Bigr|_{\Lo(\kappa),\textrm{ late }\eta'} =\,\frac{\tilde{\mu}^{4-d}\kappa H^{d-2}}{8}(-H\eta)^{d-4}\;\Im\biggl[\biggl(1+(k\eta)^2+\frac{2i}{3}(k\eta)^3\biggr)\nonumber\\
&\hspace*{0.5cm}
\times\int_{-\infty}^{\eta}\frac{\der\eta'}{(-\eta')^{8-d}}\biggl(1+(-k\eta')^2+\frac{2i}{3}(-k\eta')^3\biggr)\biggr]\;I[a(\eta)\Lambda]
\nonumber\\
&=\, \frac{\kappa H^2}{8k^3}(-H\eta)^{-2\ve}\biggl(-\frac{H^2\eta}{\tilde{\mu}}\biggr)^{\!-2\ve}\biggl[-\frac{1}{3\ve}+\frac{2}{9}-\frac{4\ve}{27}\biggr]H^{2\ve}I[a(\eta)\Lambda]\,.
\label{pwrlatet}
\end{align}
The left-over factor $H^{2\ve}$ is compensated by $\Lambda^{-2\ve}$ from $I[a(\eta)\Lambda]$.
As in the case of the trispectrum, the late-time region of the time integral has a UV-pole, which generates the time-dependent part of the logarithms of the time integral in \eqref{modexpr} in the limit $\eta\rightarrow0$.
An analogous computation for the late-time region of the mass-counterterm contribution yields
\begin{equation}
\cor{}{\phi(t,\vec k)\phi(t,-\vec k)}'\Bigr|_{\delta m^2,\textrm{ late }\eta'}=\frac{\delta m^2}{3k^3}(-\nu\eta)^{-2\delta}\biggl[-\frac{1}{2\delta}+\frac{1}{3}\biggr]\,,
\label{ctlatet}
\end{equation}
and the necessity of the analytic regulator is evident. 

In the early-time region, one obtains for the integral
\begin{align}
&\cor{}{\phi(\eta,\vec k)\phi(\eta,-\vec k)}'\Bigr|_{\Lo(\kappa),\textrm{ early }\eta'}\nonumber\\
&=\,-\frac{\tilde{\mu}^{4-d}\kappa H^{d-2}}{8}(-H\eta)^{d-4}\Im\biggl[\int_{-\infty}^0\frac{\der\eta'}{(-\eta')^{8-d}}\frac{e^{2ik\eta'}(i+k\eta')^2}{k^6}\biggr]I[a(\eta)\Lambda]
\nonumber\\
&=\,\frac{\kappa H^2}{8k^3}(-H\eta)^{-2\ve}\biggl(\frac{2e^{\gamma_E}k\tilde{\mu}}{H^2}\biggr)^{\!2\ve}\biggl[\frac{1}{3\ve}-\frac{14}{9}+\ve\biggl(\frac{100}{27}-\frac{\pi^2}{18}\biggr)\biggr]H^{2\ve}I[a(\eta)\Lambda]\,.
\end{align}
A comparison of the large square bracket with the one in~\eqref{pwrlatet} already shows that the pole terms cancel. 
Due to the different $\varepsilon$-dependent prefactors, however, the final result contains a logarithm, where the time-dependent part stems from the late-time region, while the time-independent part originates from the early-time region. 
Performing the same steps for the mass counterterm, one obtains
\begin{equation}
\cor{}{\phi(\eta,\vec k)\phi(\eta,-\vec k)}'\Bigr|_{\delta m^2,\textrm{ early }\eta}=\frac{\delta m^2}{3k^3}(-H\eta)^{-2\ve}\biggl(\frac{2e^{\gamma_E}k}{\nu}\biggr)^{\!2\delta}\biggl[\frac{1}{2\delta}-\frac{7}{3}\biggr]\,,
\end{equation}
and also here the $\delta$-pole cancels with the one in~\eqref{ctlatet}.

\section[Computation of the poles of the one-loop bispectrum of the composite operator \texorpdfstring{$\phi^2$}{Φ\^{}2}]{Computation of the poles of the one-loop bispectrum of the composite operator $\phi^2$}
\label{app::poles}

In this appendix we present the details of the computation of the pole terms of the correlator $\cor{}{\phi^2(\eta,\vec q)\phi(\eta,\vec k_1)\phi(\eta,\vec k_2)}$.

\subsection{UV-pole}
The UV-pole is obtained from the expansion of the integrand in the limit $l\gg k_{1,2}$, as explained in the main text. One therefore has to evaluate
\begin{align}
&\cor{}{\phi^2(\eta,\vec q)\phi(\eta,\vec k_1)\phi(\eta,\vec k_2)}'\Bigr|_{\textrm{UV-pole}}\nonumber\\
&=\frac{\mu^{4-d}\kappa H^d(-H\eta)^{2d-8}}{8(k_1k_2)^3}\Im\biggl[\int_{-\infty}^{\eta}\frac{\der\eta'}{(-\eta')^{8-d}}\int\frac{\der^{d-1}l}{(2\pi)^{d-1}}\frac{e^{2il(\eta'-\eta)}}{l^6}(-i+l\eta)^2(i+l\eta')^2\nonumber\\
&\hspace{0.4cm}\times\prod_{j=1}^2(-i+k_j\eta)(i+k_j\eta')\biggr|_{l\gg k_{1,2}}\biggr]\,.
\label{eq::22UVpolestart}
\end{align}
Since the time integral is finite, and we are only interested in the UV-pole, we set $d\rightarrow4$ in the integrand. Furthermore, we expand the result in the limit $l\rightarrow\infty$, keeping only the $\Lo(\eta^0,l^3)$-piece, since this is the term that gives rise to the logarithmic UV-divergence. This results in
\begin{equation}
\lim\limits_{l\rightarrow\infty}\int_{-\infty}^{\eta}\frac{\der\eta'}{(-\eta')^4}e^{2il(\eta'-\eta)}(-i+l\eta)^2(i+l\eta')^2\prod_{j=1}^2(-i+k_j\eta)(i+k_j\eta')\biggr|_{ \Lo(\eta^0,l^3)}=-\frac{il^3}{2}\,,
\end{equation}
and the expression~\eqref{eq::22UVpolestart} reduces to 
\begin{equation}
\cor{}{\phi^2(\eta,\vec q)\phi(\eta,\vec k_1)\phi(\eta,\vec k_2)}'\Bigr|_{\textrm{UV-pole}}=-\frac{\mu^{4-d}\kappa H^d(-H\eta)^{2d-8}}{16(k_1k_2)^3}\int\frac{\der^{d-1}l}{(2\pi)^{d-1}}\frac{1}{l^3}\,.
\end{equation}
To extract its UV-pole, introduce an additional IR cutoff as done in the case of the one-loop power spectrum~\eqref{IRmod}, and evaluate
\begin{equation}
\int\frac{\der^{d-1}l}{(2\pi)^{d-1}}\frac{1}{[l^2+\Lambda^2]^{\frac{3}{2}}}=\frac{\Gamma(\ve)}{4\pi^2}\biggl(\frac{4\pi}{\Lambda^2}\biggr)^{\!\ve}\,.
\label{tadpole}
\end{equation}
The expansion in $\ve$ yields the UV pole
\begin{equation}
\cor{}{\phi^2(\eta,\vec q)\phi(\eta,\vec k_1)\phi(\eta,\vec k_2)}'\Bigr|_{\textrm{UV-pole}}=\frac{\kappa H^4}{64\pi^2(k_1k_2)^3}\biggl(-\frac{1}{\ve_{\textrm{UV}}}\biggr)\,.
\label{full22UV}
\end{equation}

\subsection{IR-pole}
To extract the IR-pole, we consider the expansion of the integral in the limit $l\ll k_{1,2}$, and evaluate
\begin{align}
&\cor{}{\phi^2(\eta,\vec q)\phi(\eta,\vec k_1)\phi(\eta,\vec k_2)}'\Bigr|_{\textrm{IR-pole}}\nonumber\\
&=\frac{\mu^{4-d}\kappa H^d(-H\eta)^{2d-8}}{4(k_1k_2)^3}\Im\biggl[\int_{-\infty}^{\eta}\frac{\der\eta'}{(-\eta')^{8-d}}\int\frac{\der^{d-1}l}{(2\pi)^{d-1}}\frac{e^{i(k_1+k_2+q)(\eta'-\eta)}}{l^3q^3}(-i+l\eta)(i+l\eta')\nonumber\\
&\hspace{0.4cm}\times(-i+q\eta)(i+q\eta')\prod_{j=1}^2(-i+k_j\eta)(i+k_j\eta')\biggr]\,.
\end{align}
The time- and momentum integrals factorise, since the $\vec l$-dependence in the exponential disappears. The time integral is finite and we can let $d\rightarrow4$ in the integrand, since we are only interested in the pole term. Furthermore, we keep only the $\Lo(l^0,\eta^0)$-piece, as this is the one that generates the logarithmic IR-divergence after integrating over $\vec l$. We find
\begin{align}
&\Im\biggl[\int_{-\infty}^{\eta}\frac{\der \eta'}{(-\eta')^4}\;e^{i(k_1+k_2+q)(\eta'-\eta)}(-i+l\eta)(i+l\eta')(-i+q\eta)(i+q\eta')\nonumber\\
&\hspace{0.4cm}\times\prod_{j=1}^2(-i+k_j\eta)(i+k_j\eta')\biggr]\biggr|_{\Lo(l^0,\eta^0)}\nonumber\\
&=\frac{1}{3}\biggl[\Bigl[k^3_1+k^3_2+q^3\Bigr]\Bigl[\ln\Bigl(-e^{\gamma_E}(k_1+k_2+q)\eta\Bigr)-1\Bigr]+q^3f(k_1,k_2,q)\biggr]\,,
\end{align}
where 
\begin{equation}
    f(k_1,k_2,q)\equiv\frac{4k_1k_2}{q^2}-\frac{(k_1+k_2+q)[k_1(k_2+q)+k_2q]}{q^3}\,.
\end{equation}
As above, the momentum integral reduces to a scaleless integral. To extract its IR-pole, we use the fact that the UV- and IR-poles are the same up to their sign, so we can employ the same IR-regularised integral \eqref{tadpole} computed above, flipping the sign of the pole term obtained after expanding it in $\ve$, which then corresponds to its IR-pole. Combining these results, one obtains
\begin{align}
&\cor{}{\phi^2(\eta,\vec q)\phi(\eta,\vec k_1)\phi(\eta,\vec k_2)}'\Bigr|_{\textrm{IR-pole}}\\
&=\frac{\kappa H^4}{48\pi^2(k_1k_2)^3}\,\frac{1}{\ve_{\textrm{IR}}}\,\biggl\{\biggl[1+\frac{k^3_1+k^3_2}{q^3}\biggr]\biggl[1-\ln\Bigl(-e^{\gamma_E}(k_1+k_2+q)\eta\Bigr)\biggr]-f(k_1,k_2,q)\biggr\}\,.\nonumber
\label{full22IR}
\end{align}

\section[Computation of the integral \texorpdfstring{$I(a)$}{I(a)}]{Computation of the integral $I(a)$}
\label{app::int}

In this appendix we give the details of the computation of the ``master integral"
\begin{equation}
I(a)\equiv\int_{-\infty}^0\frac{\der\eta'}{(-\eta')^{8-d+a}}e^{iK\eta'}\int\frac{\der^{d-1}l}{(2\pi)^{d-1}}\frac{e^{il\eta'}(i+l\eta')e^{i|\vec l-\vec q|\eta'}(i+|\vec l-\vec q|\eta')}{l^3|\vec l-\vec q|^3}\,,
\label{eq::masterintstart}
\end{equation}
where $K\equiv k_1+k_2$, needed for the computation of the early-time, soft-momentum region. 

The $\vec l$-integral is computed by observing that it is the convolution of the function
\begin{equation}
f(\eta',\vec l)\equiv\frac{e^{il\eta'}(i+l\eta')}{l^3}
\end{equation}
with itself. We can thus make use of the convolution theorem~\cite{Green:2020txs}
\begin{equation}
\int\frac{\der^{d-1}l}{(2\pi)^{d-1}}f(\eta',\vec l)f(\eta',\vec q-\vec l)=\int\der^{d-1}x\;e^{i\vec q\cdot\vec x}f(\eta',\vec x)f(\eta',-\vec x)
\label{eq:selfconvolution}
\end{equation}
to evaluate this integral, where $f(\eta',\vec x)$ denotes the Fourier-transform of $f(\eta',\vec l)$ with respect to $\vec l$.
As a first step we compute
\begin{align}
f(\eta',\vec x)&=\int\frac{\der^{d-1}p}{(2\pi)^{d-1}}\;e^{i\vec p\cdot\vec x}\frac{e^{ip\eta'}(i+p\eta')}{p^3}\nonumber\\
&=\frac{\Omega_{d-3}}{(2\pi)^{d-1}}\int_{-1}^1\der\cos\theta\;(1-\cos^2\theta)^{\frac{d-4}{2}}\int_0^{\infty}\der p\;p^{d-2}e^{ipx\cos\theta}\frac{e^{ip\eta'}(i+p\eta')}{p^3}\,,
\end{align}
where the polar angle is defined as $\theta\equiv\angle(\vec p,\vec x)$ and the $d-3$ trivial angular integrations are evaluated in terms of the $d$-dimensional solid angle
\begin{equation}
    \Omega_d\equiv\frac{2\pi^{\frac{d+1}{2}}}{\Gamma(\frac{d+1}{2})}\,.
\end{equation}
The $\cos\theta$-integral gives
\begin{equation}
\int_{-1}^1\der\cos\theta\;(1-\cos^2\theta)^{\frac{d-4}{2}}e^{ipx\cos\theta}=2^{\frac{d-3}{2}}\sqrt{\pi}(px)^{\frac{3-d}{2}}\Gamma\biggl(\frac{d}{2}-1\biggr)J_{\frac{d-3}{2}}(px)\,,
\end{equation}
with $J_{\nu}(z)$ the Bessel function of the first kind. With this result, one finds
\begin{align}
f(\eta',\vec x)&=\frac{\Omega_{d-3}}{(2\pi)^{d-1}}2^{\frac{d-3}{2}}\sqrt{\pi}x^{\frac{3-d}{2}}\Gamma\biggl(\frac{d}{2}-1\biggr)\int_0^{\infty}\der p\;p^{\frac{d-7}{2}}e^{ip\eta'}(i+p\eta')J_{\frac{d-3}{2}}(px)\nonumber\\
&=\frac{i}{4\pi^{\frac{d}{2}}}\Gamma\biggl(\frac{d}{2}-2\biggr)(x^2-{\eta'}^2)^{2-\frac{d}{2}}\,.
\end{align}

Alternatively, one can Wick-rotate $t=i\eta^\prime$ and use the dispersive representation
\begin{equation}
    \frac{e^{pt}(1-pt)}{p^3} = 4\int\frac{\der\omega}{2\pi}e^{i\omega t}\frac{1}{(\omega^2+p^2)^2}
\end{equation}
to compute the Fourier transformation as
\begin{align}
   f(\eta^\prime,\vec x) &= \int\frac{\der^{d-1}p}{(2\pi)^{d-1}}e^{i\vec p \cdot \vec x}\frac{e^{ip\eta^\prime}(i+p\eta')}{p^3}=i\int\frac{\der^{d-1}p}{(2\pi)^{d-1}}e^{i\vec p\cdot\vec x + pt}\frac{(1-pt)}{p^3}\nonumber\\
   &= 4i\int\frac{\der^{d}P}{(2\pi)^{d}}e^{iP\cdot X}\frac{1}{P^4} 
   = \frac{i}{4\pi^{\frac{d}{2}}}\Gamma\biggl(\frac{d}{2}-2\biggr) \lvert X\rvert^{4-d}\,,
\end{align}
where $P\equiv(\omega, \vec p)$, $X \equiv (t, \vec x)$. Employing this in \eqref{eq:selfconvolution}, one obtains
\begin{align}
&\int\der^{d-1}x\;e^{i\vec q\cdot\vec x}f(\eta',\vec x)f(\eta',-\vec x)\nonumber\\
&=-\frac{e^{-2\pi id}}{16\pi^d}\Gamma\biggl(\frac{d}{2}-2\biggr)^{\!2}\Omega_{d-3}\int_{-1}^1\der\cos\theta\;(1-\cos^2\theta)^{\frac{d-4}{2}}\int_0^{\infty}\der x\;e^{iqx\cos\theta}x^{d-2}(x^2-{\eta'}^2)^{4-d}\nonumber\\
&=(1-e^{-i\pi d}) (2 \pi )^{-\frac{d+1}{2}} \Gamma (5-d) \Gamma \biggl(\frac{d}{2}-2\biggr)^{\!2} \biggl(-\frac{q}{\eta
   }\biggr)^{\!\frac{d-7}{2}} \Bigl[i e^{\frac{i \pi  d}{2}} J_{\frac{7-d}{2}}(q \eta )-J_{\frac{d-7}{2}}(q \eta )\Bigr]\,.
\end{align}
Hence, 
\begin{align}
    &\int\frac{\der^{d-1}l}{(2\pi)^{d-1}}\frac{e^{il\eta'}(i+l\eta')e^{i|\vec l-\vec q|\eta'}(i+|\vec l-\vec q|\eta')}{l^3|\vec l-\vec q|^3}\\
    &=(1-e^{-i\pi d}) (2 \pi )^{-\frac{d+1}{2}} \Gamma (5-d) \Gamma \left(\frac{d}{2}-2\right)^2 \left(-\frac{q}{\eta'
   }\right)^{\frac{d-7}{2}} \Bigl[i e^{\frac{i \pi  d}{2}} J_{\frac{7-d}{2}}(q \eta' )-J_{\frac{d-7}{2}}(q \eta' )\Bigr]\,.\nonumber
\end{align}
Inserting this expression into \eqref{eq::masterintstart} and evaluating the $\eta'$-integral, we find
\begin{align}
I(a)&=\frac{e^{\frac{i\pi a}{2}}(e^{2i\pi\ve}-1)\Gamma(-\ve)}{4^{1-\ve}\pi^{\frac{5}{2}-\ve}}K^a\biggl[\frac{2\Gamma (1+2\ve) \Gamma (-3-a-2\ve)\Gamma(-\ve)}{\Gamma(-\frac{1}{2}-\ve)}\biggl(\frac{K}{q}\biggr)^{\!3+2\ve}\nonumber\\
   &\quad\times{_2F_1}\left(-\frac{3+a}{2}-\ve,-1-\frac{a}{2}-\ve;-\frac{1}{2}-\ve;\frac{q^2}{K^2}\right)\nonumber\\
   &\quad+\frac{\sqrt{\pi}\Gamma (-a)\csc (\pi\ve)}{(3+2\ve) (1+2\ve)}
   {_2F_1}\left(\frac{1-a}{2},-\frac{a}{2};\frac{5}{2}+\ve;\frac{q^2}{K^2}\right)\biggr]\,,
\end{align}
where we already substituted $d=4-2\ve$. For the above computation, we need $\Im[I(2\delta)]$ and 
used the \texttt{Mathematica} package \texttt{HypExp 2.0}~\cite{Huber:2007dx} to expand first in $\delta$ and then in $\ve$. The result reads 
\begin{align}
\Im[I(2\delta)]&=\frac{(4\pi e^{\gamma_E})^{\ve}(2e^{\gamma_E}q)^{2\delta}}{12\pi^2}\biggl\{-\frac{1}{\ve^2}\frac{K[K^2-3q^2]}{q^3}+\frac{1}{\delta}\biggl[\frac{1}{\ve}-\frac{8}{3}\biggr]\nonumber\\
&\quad+\frac{1}{\ve}\biggl[-\frac{(K+q)^2(K-2q)}{q^3}\ln\biggl(\frac{K+q}{2q}\biggr)-\frac{8}{3}-\frac{7K}{2q}+\frac{K^2}{q^2}+\frac{5K^3}{6q^3}\biggr]\nonumber\\
&\quad-\frac{\pi^2}{24}\biggl[28+\frac{K^3}{q^3}-\frac{3K}{q}\biggr]+\frac{104}{9}+\frac{41K}{6q}-\frac{11K^2}{3q^2}-\frac{19K^3}{18q^3}\nonumber\\
&\quad-\biggl[\frac{16}{3}+\frac{7K}{q}-\frac{5K^3}{3q^3}\biggr]\ln\biggl(\frac{K+q}{2q}\biggr)\nonumber\\
&\quad-\frac{(K+q)^2(K-2q)}{2q^3}\ln^2\biggl(\frac{K+q}{2q}\biggr)-\frac{(K-q)^2(K+2q)}{2q^3}\ln\biggl(\frac{K}{q}\biggr)\ln\biggl(\frac{(K-q)^2}{4Kq}\biggr)\nonumber\\
&\quad+\frac{(K-q)^2(K+2q)}{q^3}\biggl[\Li_2\biggl(\frac{q}{K}\biggr)-\Li_2\biggl(-\frac{q}{K}\biggr)+\Li_2\biggl(\frac{K-q}{2K}\biggr)\biggr]\biggr\}\,.\label{eq:App:I2delta}
\end{align}

\bibliography{Bibliography}{}
\end{document}